\documentclass[twocolumn,english,showpacs,floatfix,amsmath,amssymb,prb,
superscriptaddress] {revtex4}
\usepackage[T1]{fontenc}
\usepackage[latin9]{inputenc}
\usepackage{babel}

\usepackage{graphicx}
\usepackage{amssymb}
\usepackage{amsmath}
\usepackage{esint}



\begin{document}

\newcommand{\be}{\begin{equation}}
\newcommand{\ee}{  \end{equation}}
\newcommand{\ba}{\begin{eqnarray}}
\newcommand{\ea}{  \end{eqnarray}}
\newcommand{\ve}{\varepsilon}

\title{Current-induced switching in transport through anisotropic magnetic molecules}

\author{Niels Bode} 
\affiliation{Dahlem Center for Complex Quantum Systems and Fachbereich Physik, Freie Universit\"at Berlin, 14195 Berlin, Germany}

\author{Liliana Arrachea} 
\affiliation{Departamento de Fisica, Facultad de Ciencias Exactas y Naturales, Universidad de Buenos Aires, Pab.\ I, Ciudad Universitaria, 1428 Buenos Aires, Argentina }

\author{Gustavo S.\ Lozano}
\affiliation{Departamento de Fisica, Facultad de Ciencias Exactas y Naturales, Universidad de Buenos Aires, Pab.\ I, Ciudad Universitaria, 1428 Buenos Aires, Argentina }

\author{Tamara S. Nunner}
\affiliation{Dahlem Center for Complex Quantum Systems and Fachbereich Physik, Freie Universit\"at Berlin, 14195 Berlin, Germany}

\author{Felix von Oppen}
\affiliation{Dahlem Center for Complex Quantum Systems and Fachbereich Physik, Freie Universit\"at Berlin, 14195 Berlin, Germany}

\date{\today}
\begin{abstract}
Anisotropic single-molecule magnets may be thought of as molecular switches,
with possible applications to molecular spintronics. In this paper, we consider
current-induced switching in single-molecule junctions containing an anisotropic
magnetic molecule. We assume that the carriers interact with the magnetic
molecule through the exchange interaction and focus on the regime of high currents in which the
molecular spin dynamics is slow compared to the time which the electrons spend on the molecule. In
this limit, the molecular spin obeys a non-equilibrium Langevin equation which
takes the form of a generalized Landau-Lifshitz-Gilbert equation and which we
derive microscopically by means of a non-equilibrium Born-Oppenheimer
approximation. We exploit this Langevin equation to identify the relevant
switching mechanisms and to derive
the current-induced switching rates. As a byproduct, we also derive S-matrix
expressions for the various torques entering into the Landau-Lifshitz-Gilbert equation which generalize previous expressions in the
literature to non-equilibrium situations.
\end{abstract}
\maketitle

\section{Introduction}

In recent years, electronic transport through nanostructures has witnessed a
shift toward molecular systems. Several ingenious schemes for measuring
transport through single molecules have been realized and experimental control
over such systems is rapidly improving. \cite{mol} A prime difference
between
transport
through single molecules or nanoelectromechanical systems (NEMS)
\cite{nems} as opposed to transport through more conventional nanostructures
lies in the coupling of the electronic degrees of freedom responsible for
transport to few well-defined collective modes of the molecule, with recent
research focusing on effects of molecular vibrations (molecular
nanoelectromechanics)\cite{nitzan,molvib,molvibth} and local magnetic moments
(molecular
spintronics).\cite{zuti-2004,fert-2008,bogani, fried,revs} 

The interesting property of transport setups based on single-molecule magnets is the possibility of combining some 
classical properties of macroscopic magnets with quantum features such as quantum tunneling. Work on molecular spintronics has focused on single molecule magnets such as
Mn$_{12}$ and transition metal complexes. Transport experiments with
Mn$_{12}$ concentrated on signatures of the magnetic excitations, as revealed by peaks in the
differential conductance,\cite{jo} and a spin-blockade mechanism.\cite{mn12}
Research on transition metal complexes, based e.g.\ on Co, also addresses the
Kondo effect.\cite{kondo} Related phenomena have been discussed in  molecular spin valves,
which have been realized in setups with C$_{60}$,\cite{spin-valv} and more recently
in TbPc$_2$ setups coupled to nanotubes through supramolecular interactions . \cite{supspin-valv}

In addition to their remarkable fundamental quantum transport properties, single molecule magnets are also appealing for their potential as memory cells in spintronics.\cite{memcel} In this context it
is important to have reliable mechanisms for
  writing and reading the stored information. Specifically, it is essential to
have protocols for manipulating and for detecting the orientation of the
magnetic moment.
To this end it is convenient to take advantage of the coupling between the spin
of the electrons, which tunnel from the electrodes, and the localized magnetic moment
of the molecule.\cite{timm}

Much of the existing literature, both on molecular nanoelectromechanics and
molecular spintronics, assumes that the electrons reside on the molecule for
times large compared to typical vibrational or magnetic precession periods. In
this limit, it is often appropriate to treat the dynamics of the system within a
rate- or Master equation in the exact eigenstate basis of the isolated
molecule,\cite{nitzan, ensslin} describing also spin-transfer torques out-of equilibrium.\cite{delgado-2010,delgado-2010b,molmagther}
In the context of nanoelectromechanics, there has recently been much
interest in the opposite regime of adiabatic vibrational dynamics, in which the
electronic processes are fast compared to the vibrational degrees of freedom,
e.g., in the context of certain molecular switches,\cite{mozyrsky-2006,pistolesi-2008} NEMS
near continuous mechanical
instabilities,\cite{weick-2010,weick-2011} flexural modes of suspended carbon
nanotubes and
graphene,\cite{Steele09} or the cooling and amplification of mechanical motion
by the backaction of conduction electrons.\cite{naik, stettenheim,dundas,lu,bode-2011,bode-2011b}

The goal of the present work is to explore the latter limit in the context of magnetic molecules.
We consider a generic model for the magnetic molecule, which includes an easy axis anisotropy sandwiched between two metallic (possibly polarized) electrodes at which a bias voltage is applied.\cite{fried}
We focus on the regime where the typical time for dynamics of the molecular magnetic moment is much larger than the dwell time of the electrons flowing through the structure. Within this {\em adiabatic} regime it is possible to study the coupled electronic transport and spin dynamics within a  non-equilibrium Born-Oppenheimer approximation (NEBO) analogous to the one adopted in NEMS in the equivalent regime.\cite{mozyrsky-2006,pistolesi-2008}
Starting from a microscopic description, we can derive semiclassical equations of motion for the local magnetic moment that have the structure of generalized Landau-Lifshitz-Gilbert (LLG) equations. The latter have been the basis for several previous works in spintronics and nanomagnetism.\cite{Katsura06,kupferschmidt-2006,fransson-2008,nunez-2008,basko-2009,dunn-2011, lopez-monis-2012,tserkovnyak-2002,brataas-2008,brataas-2011,hals-2010} We note that in previous works describing magnetic nanoparticles, LLG equations have been derived in a perturbative way assuming either that the coupling between the
electronic spin and the magnetic moment of the nanoparticle is small
\cite{Katsura06,fransson-2008,nunez-2008} and/or that tunneling between the leads and nanoparticle
is weak.\cite{Katsura06,fransson-2008,lopez-monis-2012} In contrast, our microscopic derivation relies entirely on the non-equilibrium
Born Oppenheimer approximation which is valid in the high-current limit as
described above. As a consequence our non-perturbative approach allows us to compute how the parameters of the LLG equation depend on the state of the molecular moment as well as on the applied bias and gate voltages.
We mainly focus on two important features. First we analyze the magnetic molecule attached to (non-magnetic) metallic leads. In this case, switching of the molecular  moment is induced by the fluctuating torque exerted by the current flow. In addition, we also investigate the renormalization of the switching barrier by the average torque caused by the charge carriers. Second, we consider that switching is dominated by a different mechanism for spin-polarized electrodes, namely by the spin-transfer torque exerted by the transport current. 
The latter is well known in the context of layered magnetic structures.\cite{slonczewski-1996,berger-1996,spin-torque} We also analyze the behavior of the electronic current and we identify in this quantity the interplay between the spin fluctuations and the signatures of coherent transport, which are typical of the molecular devices.

This paper is organized as follows. In Sec.\ \ref{sec:Model}, we introduce our
model of the single-molecule junction containing an anisotropic magnetic
molecule. The Landau-Lifshitz-Gilbert equation describing the dynamics of the
local moment of
the molecule is derived within the NEBO approximation in Sec.\
\ref{sec:Derivation} and related to scattering matrix theory in Sec.\ \ref{sec:smat}.
This Langevin equation is explored in Sec.\ \ref{sec:applications}. Switching
of the molecular moment is discussed in Secs.\ \ref{sec:Fluctuation} and
\ref{sec:Torque}. Section
\ref{sec:Fluctuation} focuses on switching caused by fluctuations while
Sec.\ \ref{sec:Torque} discusses situations when the switching is dominated by
the spin-transfer torque. We conclude in Sec.\ \ref{sec:Conclusion}. Some
technical details are relegated to appendices.

\section{Model}
\label{sec:Model}

We consider a minimal model of an anisotropic magnetic molecule embedded into a
single-molecule junction.\cite{nazarov} We assume that transport through the
molecule is dominated by a single molecular orbital which is coupled to left
($L$) and right ($R$) leads at different chemical potentials. The spin
{\boldmath ${\bf \hat s}$} of the current-carrying electrons couples to a
localized molecular spin $\hat{\bf M}$ through exchange. Then, the full
Hamiltonian 
\be
H= \sum_{\alpha=L,R}( H_{\alpha} + H_{d \alpha}) + H_d, \label{eq:Hamiltonian}
\ee
encompasses the Hamiltonians
\be
H_{\alpha}=\sum_{k_{\alpha},\sigma} \epsilon_{k_{\alpha},\sigma}
c^{\dagger}_{k_{\alpha},\sigma}c_{k_{\alpha},\sigma}
\ee
of the left ($L$) and right ($R$) leads, modeled as free-electron systems
(creation operators $c^\dagger_{k_{\alpha},\sigma}$). We will consider the
possibility of spin-polarized leads, assuming a spin-dependent
dispersion $\epsilon_{k_\alpha,\sigma}$. The tunneling Hamiltonian
\be
  H_{d \alpha}= \sum_{k_{\alpha},\sigma} w_{k_{\alpha}} 
(c^{\dagger}_{k_{\alpha},\sigma} d_{\sigma}+d^{\dagger}_{\sigma} c_{k_{\alpha},\sigma})
\ee
describes the hybridization between the molecular orbital (with creation operator $d^{\dagger}_{\sigma}$) and the leads. The molecular Hamiltonian is given by
\ba
H_d &= & \sum_{\sigma} \epsilon_0 d^{\dagger}_{\sigma} d_{\sigma} + g_e\hat{\bf s}\cdot{\bf B} + J \hat{{\bf s}} \cdot \hat{\bf M} + U(\hat{\bf M}).
\label{dotHamiltonian}
\ea
The potential experienced by the molecular spin in the absence of coupling to the external leads is $U(\hat{\bf M}) =  g_d \hat{\bf M}\cdot{\bf B} -D {\hat M}_z^2$.
The uniaxial anisotropy of the molecule is parametrized through the anisotropy parameter $D$, with easy-axis anisotropy corresponding to $D>0$ and easy-plane anisotropy to $D<0$.
The coupling constant $J$ denotes the strength of the exchange interaction between the molecular spin ${\bf \hat M}$ and the electronic spins,
\be
\hat{s}_j = \frac{\hbar}{2}
\sum_{\sigma,\sigma'} d^{\dagger}_{\sigma} \sigma^{j}_{\sigma \sigma^{\prime}} 
d_{\sigma^{\prime}},
\ee
where $\sigma^{j}$ (with $j=x,y,z$) are the Pauli matrices. For simplicity, we assume this exchange interaction to be isotropic. The energy of the molecular orbital $\epsilon_0$ can be tuned by a gate voltage and ${\bf B}$ represents a Zeeman field acting on the electronic and the localized spins with $g$-factors $g_e$ and $g_d$, respectively.

\section{Description of the spin dynamics}
\label{sec:Derivation}
We now discuss this model in the limit of slow precession of the magnetic
moment, that is, many electrons are passing the molecule during a single
precessional period of the molecular spin. In this limit, it is natural to
approximate the molecular spin as a classical variable whose dynamics can be
described within a non-equilibrium Born-Oppenheimer approximation. The resulting
dynamics takes the form of a Langevin equation of the Landau-Lifshitz-Gilbert
type which we derive microscopically for our model. Specifically, the exchange
coupling between the current-carrying electrons and the molecular moment
introduces additional torques and damping terms which enter into the Langevin
equation and which we will now discuss in detail.

\subsection{Semiclassical equation of motion of the molecular spin}  

Our derivation starts from the Heisenberg equation of motion for the molecular spin,
\begin{eqnarray}
\dot{\hat{ M}}_j &=&  \sum_{l,k} \varepsilon_{j l k}[J  \hat{s}_l + g_d B_l ] \hat{M}_k 
\nonumber\\
&&+
D \sum_k \varepsilon_{z j k} [ \hat{M}_z \hat{M}_k+ \hat{M}_k \hat{M}_z],
\end{eqnarray}
where $\varepsilon_{jlk}$ is the antisymmetric Levy-Civita tensor. Within the non-equilibrium Born-Oppenheimer approximation, we can turn this into an equation of motion for the expectation value ${\bf M}(t)= \langle \hat{\bf M}(t) \rangle$ of the localized spin, 
\begin{eqnarray} 
\label{eqmot}
\dot{\mathbf{M}}= \mathbf{M} \times\left[-\partial_{\mathbf{M}} U(\mathbf{M}) -
 J \mathbf{s} + \boldsymbol{\delta {\cal B}}\right],
\end{eqnarray}
with $-\partial_{\mathbf{M}} U(\mathbf{M})=- g_d \mathbf{B} + 2 D M_z \hat{\mathbf{e}}_z$.
Here, ${\bf M}={\bf M}(t)$ denotes the molecular spin averaged over a time
interval large compared to the electronic time scales, but small compared to the
precessional dynamics of the molecular spin itself. The corresponding
time-averaged electronic spin ${\bf s}={\bf s}(t)$ can be expressed in terms of
the 
electronic lesser Green's function
\be 
{G}_{\sigma \sigma^{\prime}}^<(t,t^{\prime}) = 
i \langle d^{\dagger}_{\sigma^{\prime}} (t^{\prime}) d_{\sigma}(t)\rangle
\ee 
of the molecular orbital as
\be \label{s}
s_j(t)=\langle \hat{s}_j\rangle_{{\bf M}(t)} = -\frac{i\hbar}{2} \mbox{tr} [ G^<(t,t) {\sigma}^{j}].
\ee
It is important to note that due to the Born-Oppenheimer approximation, the
lesser Green's function must be evaluated for a given time dependence of the
molecular spin ${\bf M}(t)$. As a result, the average electronic spin ${\bf
s}(t)$ depends on the molecular spin at earlier times. This will be considered
in more detail in the next subsection. The instantaneous contribution gives rise
to a force acting on the molecular spin. Retardation effects produce terms
proportional to $\dot{\mathbf{M}}$, appearing in the equation of motion as
Gilbert damping and a change in the gyromagnetic ratio. Additionally,
fluctuations of the electron spin give rise to a fluctuating Zeeman field
$\boldsymbol{\delta {\cal B}}$ acting on the molecular spin.

\subsection{Electronic Green's function in the adiabatic limit}
We now turn to evaluate the electronic lesser Green's function, accounting for
the slowly varying molecular spin ${\bf M}(t)$. We start by considering the
corresponding retarded Green's function
\be 
{G}^R_{\sigma, \sigma^{\prime}}(t,t^{\prime}) = 
-i \theta(t-t') \langle \{d_{\sigma} (t), d^{\dagger}_{\sigma^{\prime}}(t^{\prime}) \}
\rangle.
\ee
Since the electrons are non-interacting, we can obtain ${G}^<$ from ${G}^R$ at
the end of the calculation.  From now on, we set $\hbar=1$. The retarded Green's function
satisfies the Dyson equation\cite{jauho}
\begin{multline} \label{dyret}
 (-i \partial_{t^{\prime}} -\epsilon_0) {G}^R_{\sigma
\sigma^\prime}(t,t^{\prime})
- \int dt_1 \sum_{\sigma_1} {G}^R_{\sigma \sigma_1}(t,t_1)
\Sigma^R_{\sigma_1\sigma^\prime}(t_1,t^{\prime})\\
-\sum_{\sigma_1}
G^R_{\sigma\sigma_1}(t,t^\prime){\boldsymbol
\sigma}_{\sigma_1\sigma^\prime}\cdot {\bf b}(t^\prime)  
=\delta(t-t^{\prime}) \delta_{\sigma \sigma^{\prime}}.
\end{multline}
Here we introduce the self-energy
\begin{align}
\Sigma^R_{\sigma,\sigma'}(t,t^{\prime}) = 
\delta_{\sigma \sigma'} \sum_{\alpha} \int \frac{d\omega}{2 \pi}
\Sigma^R_{\alpha,\sigma}(\omega) e^{-i\omega(t-t')}
\end{align}
with
\begin{align}\label{eq:selfenergyR}
\Sigma^R_{\alpha,\sigma}(\omega) =
\sum_{k_{\alpha}} |w_{k_{\alpha}}|^2  \frac{1}{\omega -
\epsilon_{k_{\alpha},\sigma} + i \eta}
\end{align}
accounting for the coupling to the (possibly spin-polarized) leads, see
also Appendix \ref{app:coefficients}. It is convenient to introduce an effective magnetic field experienced by
the electrons  given by
\begin{align}
\mathbf{b}(t)=\frac{1}{2}(J{\bf M}(t)+ g_e {\bf B}). \label{eq:fieldb1}
\end{align}
Notice that even if we consider a constant external magnetic field, the
effective magnetic field is time dependent due to
the explicit time dependence of the
molecular spin $\mathbf{M}=\mathbf{M}(t)$.

In order to implement the Born-Oppenheimer approximation, it is convenient to
rewrite the Dyson equation in the mixed (Wigner) representation defined by  
\begin{equation}
\tilde{O}(t,\omega)\equiv\int\mathrm{d}\tau\,\mathrm{e^{i\omega\tau}O(t+\tau/2,
t-\tau/2)},
\end{equation}
for a general quantity $O(t_1,t_2)$ depending on two times with central and
relative times defined as $t=(t_{1}+t_{2})/2$ and $\tau=t_{1}-t_{2}$.
The non-equilibrium Born-Oppenheimer approximation can now be implemented by
noting that the dependence on the central time $t$ is slow. Thus, convolutions
such as $C(t_{1},t_{2})=\int\mathrm{d}t_{3}\, C_1(t_1,t_{3})C_2(t_{3,}t_{2})$
can be approximated in Wigner representation through
\begin{align}
\tilde{C}(t,\omega)   &=   \exp\left[\frac{i}{2}\left(\partial_{\omega}^{\tilde
{C_1}}\partial_{t}^{\tilde {C_2}}-\partial_{t}^{\tilde{C}_1}\partial_{\omega}^
{\tilde{C}_2}\right)\right]\tilde C_1\tilde C_2
\nonumber \\
 &  \simeq  \tilde C_1 \tilde
C_2 + \frac{i}{2}\left(\partial_{\omega}\tilde C_1\partial_{t}\tilde
C_2-\partial_{t}\tilde C_1\partial_{\omega}\tilde
C_2\right)
\end{align}
in next-to-leading order using the shorthand $\tilde{C}_l=\tilde{C}_l(t,\omega)$.

For our problem, to lowest order
in the slow changes of $\bf{M}$, we then obtain for the Dyson equation
\ba
1&=& G^{R} \left(\omega-\epsilon_{0} -\Sigma^{R}
-\boldsymbol{\sigma}\cdot\mathbf{b}\right)\nonumber\\
&&-\frac{i}{2}\partial_{\omega}G^{R}\partial_{t}\boldsymbol{
\sigma}\cdot\mathbf{b}-\frac{i}{2}\partial_{t}G^{R}\left(1-
\partial_{\omega}\Sigma^{R}\right)
\label{eq:wignerconv}
\ea
where $G^{R}=G^{R}(t,\omega)$ denotes the Green's function in the Wigner
representation.
In the above equation and in what follows, the Green's functions, as well as the
self-energy, are matrices in spin space with elements  
$G^{R}_{\sigma, \sigma^{\prime}}(t,\omega)$ and
$\Sigma^R_{\sigma,\sigma^{\prime}}(\omega)$,
respectively. In the strictly adiabatic limit we drop the terms proportional to
derivatives
with respect to the central time. To this order we obtain
\begin{equation}\label{eq:G0Rinv}
G_{0}^{R}(\mathbf{M},\omega)=\left[\omega-\epsilon_{0}
-\Sigma^{R}(\omega)
-\boldsymbol{\sigma}\cdot\mathbf{b}\right]^{-1}.
\end{equation}
In next-to-leading order in the Born-Oppenheimer approximation, we keep the time
derivatives with respect to central time to linear order.
Equation \eqref{eq:G0Rinv} implies $[\partial_t G_{0}^{R}]^{-1} = - \boldsymbol{\sigma}\cdot \dot{\mathbf{b}}$. Accordingly, by differentiating $\left[\omega-\epsilon_{0}
-\Sigma^{R}(\omega)-\boldsymbol{\sigma}\cdot\mathbf{b}\right]G_{0}^{R} = 1$ with respect to time and multiplying the resulting equation with $G_{0}^{R} $ one obtains the identity $\partial_t G_{0}^{R}= G_{0}^{R} \boldsymbol{\sigma}\cdot \dot{\mathbf{b}}
G_{0}^{R}$. Then the Dyson equation yields
\ba   
G^R\simeq G_0^R +\frac{i}{2}\left[\partial_\omega G^R_0 \boldsymbol{\sigma}\cdot
\dot{\mathbf{b}} G^R_0 - G^R_0 \boldsymbol{\sigma}\cdot
\dot{\mathbf{b}}\partial_\omega G^R_0\right].
\ea

The lesser Green's function can now be deduced from the relation $G^<=G^R
\otimes \Sigma^< \otimes G^A$,\cite{jauho} where $\otimes$ denotes
integration over internal
time arguments and $G^A=[G^R]^{\dagger}$. The lesser self-energy depends only on time differences,
\begin{align} 
\Sigma_{
\sigma,\sigma^\prime}^<(\omega)= i \delta_{\sigma,\sigma^{\prime}}
\sum_{\alpha} f_{\alpha}(\omega) 
\Gamma_{\alpha,\sigma} (\omega).\label{eq:sigmaless}
\end{align} 
Here, we introduced $\Gamma_{\alpha,\sigma} (\omega)= -2
\mbox{Im}[\Sigma^R_{\alpha,\sigma}(\omega)]$ as well as the Fermi functions
$f_{\alpha}(\omega) = 1/(1+e^{\beta_{\alpha}(\omega -\mu_{\alpha})})$ with
$\alpha=L,R$. We obtain after straightforward algebra
\begin{multline}
G^<\simeq G_0^< +\frac{i}{2} \Bigl(\partial_{\omega}G_0^<
\boldsymbol{\sigma}\cdot \dot{\mathbf{b}} G_0^A - G_0^< \boldsymbol{\sigma}\cdot
\dot{\mathbf{b}} \partial_{\omega}G_0^A \\+\partial_{\omega}G_0^R
\boldsymbol{\sigma}\cdot \dot{\mathbf{b}} G_0^< - G_0^R \boldsymbol{\sigma}\cdot
\dot{\mathbf{b}} \partial_{\omega}G_0^< \Bigr).
\label{eq:Gless}
\end{multline}
Here we used $G_0^<=G_0^R\Sigma^<G_0^A$ and suppressed the arguments of the frozen
Green's functions, $G_0^{R,A,<}=G_0^{R,A,<}(\bf{M},\omega)$.

\subsection{Electron spin}
We can now employ this result for the electronic Green's function and evaluate the
electron spin. Substituting Eq.\ (\ref{eq:Gless}) into
Eq.\ (\ref{s}), we find
\begin{align} \label{eq:sapprox}
J\mathbf{s}({\bf M})&\simeq J\mathbf{s}^0({\bf M}) + \gamma({\bf M})\dot{\mathbf{M}}.
\end{align}
The first term in \eqref{eq:sapprox} contains the average electron spin
\ba 
s^0_l({\bf M}) =  -\frac{i}{2} \int \frac{d\omega}{2\pi} 
\mbox{tr}[G_0^< \sigma^l ]
\label{s0}
\ea
in the strictly adiabatic limit. The correction due to retardation effects
associated with the slow dynamics of the molecular spin are captured by the
matrix $\gamma({\bf M})$,
\begin{align}\label{eq:gamma}
\gamma^{lk}({\bf M})= &\frac{J^2}{4} \int \frac{d\omega}{2\pi}
\Bigl\{\mbox{tr}[\partial_{\omega} G_0^> \sigma^k
G_0^<
\sigma^l]\nonumber\\
  &\quad+\mbox{tr}[\partial_{\omega} G_0^A \left(\sigma^k
G_0^<
\sigma^l-\sigma^l G_0^<
\sigma^k\right)]\Bigr\},
\end{align}
where we have integrated by parts and used the greater Green's function
\begin{align}
G^{>}_{\sigma \sigma^{\prime}}(t,t^{\prime})=-i \langle d_{\sigma}(t) d_{\sigma^{\prime}}^{\dagger}(t^{\prime}) \rangle,
\end{align} 
with the relation $G^>-G^<=G^R-G^A$. It is appropriate to split this matrix into $\gamma^{lk}
=\gamma^{lk}_{s}+\gamma^{lk}_{a}$ with the shorthand
$\gamma^{lk}_{s,a}=(\gamma^{lk}\pm \gamma^{kl})/2$, see Eqs.\ \eqref{eq:gammasGr} and
\eqref{eq:gammaaGr}.
As we will see, the symmetric part $\gamma^{lk}_{s}$
describes Gilbert damping of the molecular spin, induced by the coupling to the electrons
while the  antisymmetric part $\gamma^{lk}_{a}$ will induce a coupling
renormalization.

Due to the stochastic nature of the current flow through the magnetic molecule
(as reflected in thermal as well as shot noise of the current), the electronic spin
will also fluctuate, giving rise to a fluctuating torque $\boldsymbol{\delta
{\cal B}}(t)$
acting on the molecular spin. Using Wick's theorem we obtain for the symmetrized correlator
\begin{align} 
\langle \delta B_k(t) \delta B_m(t^{\prime}) \rangle  =  \frac{J^2}{4} 
\mbox{tr}[\sigma^k G^>(t,t^{\prime}) 
\sigma^m G^<(t^{\prime},t) ]_s
\label{flucspin}
\end{align}
of the electron spin.
In the Born-Oppenheimer approximation, the fluctuations of the spin, as given by
Eq.\ (\ref{flucspin}), can be evaluated using the Green's function $G^{<,>}_0$
to lowest order in $\dot{\mathbf{b}}$. Thus, the fluctuating Zeeman field
$\boldsymbol{\delta{\cal B}}$ has the symmetrized correlator $\langle \delta{\cal B}_k(t)
\delta{\cal B}_l(t^{\prime})\rangle  =
\tilde{D}^{kl}(\mathbf{M})\delta(t-t^{\prime})$
with
\begin{align}
\tilde{D}^{kl}(\mathbf{M})=   \frac{J^2}{4} \int \frac{d \omega}{2
\pi}\mbox{tr}[\sigma^k
G_0^>
\sigma^l G_0^< ]_s.\label{eq:variance}
\end{align}
Note that in the Born-Oppenheimer limit, we can neglect any frequency dependence
of this correlation function on the time scales of the molecular spin, so that
the fluctuating Zeeman field can be taken as locally correlated in time. 

\subsection{Landau-Lifshitz-Gilbert equation}
Substituting the expression for the electronic spin  (\ref{eq:sapprox}) into the equation
of motion (\ref{eqmot}) we obtain a Langevin equation  of the Landau-Lifshitz-Gilbert type, 
\begin{align}\label{eq:LLG1}
\dot{\bf M} =  {\bf M} \times \left[-\partial_{\mathbf{M}} U - J \mathbf{s}^0 
- \gamma_s \dot{\bf M}  - \gamma_a \dot{\bf M} +\boldsymbol{\delta{
\cal B}} \right].
\end{align}
Note that, unlike in simple versions of a Landau-Lifshitz-Gilbert equation,
the effective exchange field ${\bf s}^0$ as well as the coefficient matrices
$\gamma_s$ and $\gamma_a$ still depend on the molecular spin ${\bf M}$
itself. We can simplify this equation by introducing the vector
\begin{align}\label{eq:C}
  C_k(\mathbf{M}) &= \frac12 \sum_{lm} \epsilon_{klm} \gamma^{lm}_a(\mathbf{M}).
\end{align}
Using that the length of $\mathbf{M}$ is conserved, it follows that
the antisymmetric part of $\gamma$ merely renormalizes the precession frequency by an overall prefactor
\begin{align}
  \alpha(\mathbf{M}) &= \frac{1}{1+ \mathbf{C} \cdot \mathbf{M}}.
\label{eq:prefactor}
\end{align}
This yields the simplified Landau-Lifshitz-Gilbert equation 
\begin{align}\label{eq:LLG2}
\dot{\bf M} = \alpha  {\bf M} \times \left[-\partial_{\mathbf{M}} U - J
\mathbf{s}^0  - \gamma_s \dot{\bf M} +\boldsymbol{\delta{
\cal B}} \right],
\end{align}
which we will analyze further in the subsequent sections. 

When coupled to spin-polarized leads and when a finite bias voltage is applied,
the torque can be non-conservative, yielding the so-called spin-transfer
torque.\cite{spin-torque} Also the eigenvalues of $\gamma_s$ can become overall
negative, providing another mechanism of energy transfer from the electrons to
the localized spin.

It is interesting to compare these results with those for the related problem of
charge carriers interacting with a slow vibrational degree of freedom in a NEMS.
In both cases, the dynamics of the slow collective degree of freedom can be
described in terms of a Langevin equation.\cite{mozyrsky-2006,pistolesi-2008} Since the
stochastic spin dynamics is
effectively two-dimensional, it generically exhibits similar phenomena as NEMS
with more than one vibrational mode.\cite{dundas,lu,bode-2011} Specifically,
this includes the
non-conservative nature of the average force in general non-equilibrium
situations as well as the presence of the antisymmetric contribution to the
velocity-dependent force. The latter Berry phase
contribution\cite{berry} acts, however, in different ways in the two cases,
owing to the different orders of the Langevin equation. In the vibrational
context, this term gives rise to an effective Lorentz force, while it merely
renormalizes the precession frequency in the context of the magnetic molecule.

\section{Relation to scattering matrix theory}\label{sec:smat}
Before proceeding with analyzing the Landau-Lifshitz-Gilbert equation
(\ref{eq:LLG2}) in more detail, we pause to provide S-matrix expressions for the
various entries into this equation. It has already been noted in a series of
works by Brataas \textit{et al.}\cite{tserkovnyak-2002,brataas-2008,brataas-2011,hals-2010} that the coefficients in the Landau-Lifshitz-Gilbert equation in lead-ferromagnet-lead structures can be expressed in terms of the scattering matrix of the structure, resulting in expressions for Gilbert damping and the fluctuating torque in thermal equilibrium and for current-induced spin-transfer torques within linear response theory. Here we will provide S-matrix expressions which remain valid in general out-of-equilibrium situations and which include the exchange field and the precession renormalization in addition to the Gilbert damping with the only assumption that the precessional frequency of the localized magnetic moment is slow compared to the electronic time scales. Our
discussion here closely follows recent work on current-induced forces in
nanoelectromechanical systems.\cite{bode-2011,bode-2011b}

For adiabatic parameter
variations, the full ${\cal S}$-matrix of mesoscopic conductors can be expressed
in the Wigner representation as ${\cal S}(t,\omega)= 1-2\pi i (W
{G}^R W^\dagger)(t,\omega)$. Expanding ${\cal S}$ to linear
order in the velocities $\dot{\bf M}$ of the adiabatic variables, Moskalets and
B\"uttiker \cite{moskalets-2004,arrachea-2006} introduced an $A$-matrix through
\begin{align}
{\cal S}(t,\omega) \simeq S({\bf M}(t),\omega)+ \dot{\mathbf{M}}(t) \cdot
\mathbf{A}({\bf M}(t),\omega).
\end{align}

For the model considered here, the frozen $S$-matrix
is readily related to the frozen retarded Green's function $G_0^R({\bf
M},\omega)$ through
\cite{nazarov}
\begin{equation}
  S({\bf M},\omega) = 1 - 2\pi i W G_0^R({\bf M},\omega) W^\dagger 
  \label{SMatrix}
\end{equation}
while the $A$-matrix is given by
\begin{align} \label{Amat}
A_k({\bf M},\omega) = \pi \frac{J}{2} \Bigl[&\partial_\omega \left(W
G_0^R\right) \sigma^k G_0^R W^\dagger \nonumber\\
  &\qquad- W
G_0^R \sigma^k \partial_\omega \left(G_0^R W^\dagger \right) \Bigr].
\end{align} 

The average electronic spin $s^0_l({\bf M})$ can be
written in terms of the frozen $S$-matrix \eqref{SMatrix} by expressing
the lesser Green's function $G_0^<=G_0^R\Sigma^< G_0^A$ in terms of the
self-energy
$\Sigma^<(\omega)= i \pi \sum_{\alpha} f_\alpha W^\dagger \Pi_\alpha W$ with
$\Pi_\alpha$ a projector on lead $\alpha$.
Using the identity $2\pi i W^\dagger W = (G^R)^{-1}-(G^A)^{-1}$, we then
find
\begin{equation} \label{eq:meanSmat}
 J s^0_k({\bf M}) = -\sum_\alpha \int \frac{d\omega}{2\pi i} 
f_\alpha \
{\rm Tr} \left(\Pi_\alpha S^\dagger \frac{\partial S}{\partial M_k} \right)
\end{equation}
for the average electronic spin. Here the trace
``Tr'' acts in lead-channel space. 

The S-matrix expression (\ref{eq:meanSmat})
allows us to make some general statements about the average torque acting on
the molecular spin. In particular, we can evaluate the curl of the
average torque, 
\begin{align}\label{curl}
\frac{\partial (J s^0_k)}{\partial M_{l}}-\frac{\partial (J s^0_{l})}
{\partial M_k} = i \sum_\alpha \int \frac{d\epsilon}{\pi}
f_\alpha  {\rm Tr} \left( \Pi_\alpha \frac{\partial S^\dagger}{\partial M_l}
\frac{\partial S}{\partial M_{k}} \right)_a .
\end{align}
In thermal equilibrium, Eq.~\eqref{curl}
can be turned into a trace over a commutator of finite-dimensional matrices due
to the relations $f_\alpha=f$, $\sum_\alpha \Pi_\alpha=1$,
and unitarity $S^\dagger S={\bf 1}$. This implies that $\partial
(S^\dagger S)/\partial M_k =0$ so that there is no spin-transfer torque. In
general
out-of-equilibrium situations, the curl will be nonzero, giving
rise to finite spin-transfer torque. 

Similar to the average spin, we can also
express the variance of the fluctuating Zeeman field \eqref{eq:variance} in
terms of the frozen $S$-matrix,
\begin{align}
\label{variance2}
\tilde{D}^{kl}({\bf M})=& \sum_{\alpha\alpha'} \int\frac{d\omega}{2\pi}
f_\alpha(1-f_{\alpha'})\nonumber  \\
 &\times {\rm Tr} \left\{ \Pi_\alpha \left( S^\dagger 
\frac{\partial S}{\partial M_{k}}\right)^\dagger \Pi_{\alpha'} S^\dagger 
\frac{\partial S}{\partial M_{l}} \right\}_s.
\end{align}
By going to a basis in which $\tilde{D}$ is diagonal and using
$\Pi_\alpha=\Pi_\alpha^2$, we find that $\tilde{D}$ is a positive definite
matrix. 

To express the velocity-dependent forces in terms of the scattering matrix in
general non-equilibrium situations, we need to go beyond the frozen scattering
matrix $S$ and include the $A$ matrix introduced above. The
Gilbert-damping coefficients appearing in the Langevin equation
\eqref{eq:LLG2} can then be written as
\begin{align}\nonumber
&\gamma_s^{kl}({\bf M}) = \sum_\alpha \int \frac{d\omega}{4\pi}
(-\partial_\omega
f_\alpha) {\rm Tr} \left\{ \Pi_\alpha \frac{\partial S^\dagger} {\partial M_k}
\frac{\partial S}{\partial M_{l}}\right\}_s \\ 
 \label{damp2}
 &+ \sum_{\alpha} \int \frac{d\omega}{2\pi i} f_\alpha
{\rm Tr} \left\{ \Pi_{\alpha}\left( \frac{\partial S^\dagger}{\partial M_k}
A_{l} - A^\dagger_{l} \frac{\partial S}{\partial M_k} \right)
\right\}_s .
\end{align}
The eigenvalues of the first line are strictly positive while the sign of
the second line is not fixed, giving rise to the possibility of
overall \textit{negative} Gilbert
damping.
Note that the second line is a pure non-equilibrium contribution. This can
be seen by using
unitarity of $S$ as well as ${\cal S}$, implying \cite{moskalets-2004,arrachea-2006}
${\bf A}S^\dagger + S {\bf A}^\dagger = (i/2)[\partial_{\bf M}
S\partial_\omega S^\dagger - \partial_\omega S \partial_{\bf M}
S^\dagger]$. With this preparation, it is now easy to ascertain that in
equilibrium damping and fluctuations are related by
the fluctuation-dissipation theorem, $\tilde{D}^{kl} = 2 T \gamma^{kl}_s$.

Similarly, we express the antisymmetric part of $\gamma^{kl}$ as
\begin{equation}
\gamma_a^{kl}({\bf M})= \sum_{\alpha} \int \frac{d\omega}{2\pi i} f_\alpha
{\rm Tr} \left\{ \Pi_{\alpha}\left( S^\dagger \frac{\partial A_{k}}{\partial
M_{l}} -
\frac{\partial A^\dagger_{k}}{\partial M_{l}} S\right)  \right \}_a,
\label{effB2}
\end{equation}
which causes a renormalization of the precession frequency, as discussed above.

\begin{figure}[t]
\includegraphics[width=4cm,keepaspectratio=true]{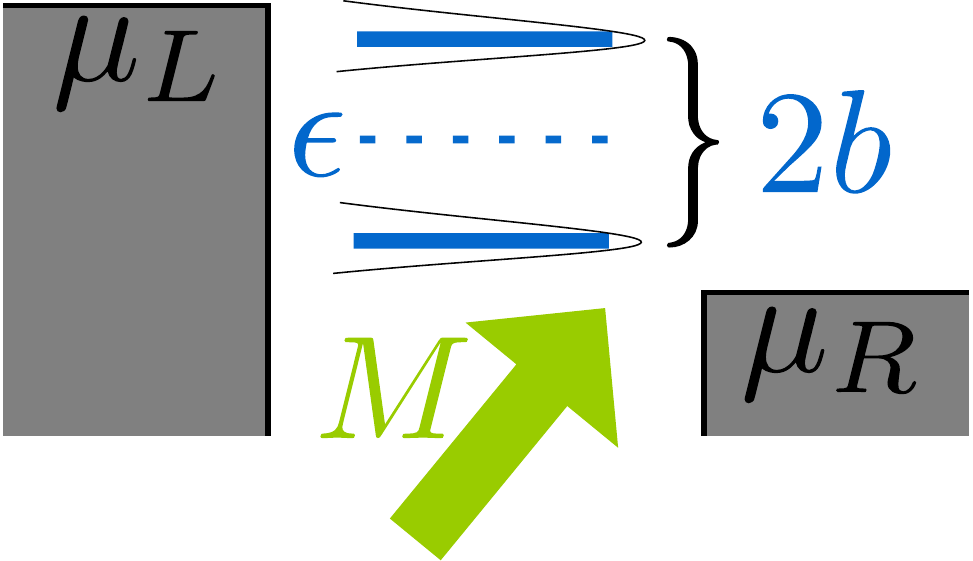}
\caption{(Color online) Illustration of the setup. The electronic
level splits due to the effective magnetic field, given by
$b=b(\vartheta)$. The number of levels in the
current window depends on $b$, the applied bias voltage $eV=\mu_L-\mu_R$ and the
gate voltage $e V_{\mathrm{g}} = (\mu_L+\mu_R)/2 -\epsilon$. It changes from
zero ($eV/2>eV_{\mathrm{g}} \pm b$) to one ($eV_{\mathrm{g}} +
b>eV/2>eV_{\mathrm{g}} - b$) to two ($eV_{\mathrm{g}} \pm b>eV/2$) [assuming
$eV,eV_{\mathrm{g}}>0$].}\label{fig:sketch01}
\end{figure}
\section{Molecular switches with axial symmetry}\label{sec:applications}
From now on we specify to the case of axial symmetry, where both the magnetic
field and the polarization of the
leads point along the anisotropy axis. In this section, we
will derive explicit expressions for the current-induced forces, including their
dependence on the molecular spin $\mathbf{M}$.

We first consider the average torque which is determined by the average
electronic spin. Given that there are two basic vectors in the problem, namely
$\hat{\bf e}_z$  and $\hat{\mathbf{M}}=\mathbf{M}/M$, the spin can be decomposed as
\begin{equation}\label{eq:s0uniaxial}
{\bf s}^0({\bf M})  =  s_M({\bf M}) \hat{{\bf M}} +  s_z({\bf M})
\hat{\bf e}_z + s_t({\bf M}) (\hat{\bf e}_z \times  \hat{{\bf M}} ).
\end{equation}
Hence, the average torque exerted on the molecular
spin by the conduction electrons is
\begin{align}
   -\alpha  {\bf M} \times  J \mathbf{s}^0  =  -\alpha  {\bf M} \times
 \left[s_z\,
\hat{\bf e}_z + s_t\, (\hat{\bf e}_z \times  \hat{{\bf
M}})\right]\label{eq:avtorque}
\end{align}
which is obtained by inserting Eq.\ \eqref{eq:s0uniaxial} into the Landau-Lifshitz-Gilbert equation \eqref{eq:LLG2}.
The first term inside the bracket can be derived from a
potential, since its curl vanishes. This becomes more
evident from the explicit expressions below using that the
$\mathbf{M}$-dependence of the coefficients stems from the effective magnetic
field $\mathbf{b}$ experienced by the electrons and that the length of
$\mathbf{M}$ is conserved. This contribution modifies the precession
frequency around the $z$-axis. In contrast, the second term on the right
hand side of Eq.\ \eqref{eq:avtorque} has a non-vanishing curl,
$\nabla_{\mathbf{M}} \times
\left[s_t\, (\hat{\bf e}_z
\times  \hat{{\bf M}})\right] \neq 0$, so that $s_t$ introduces
a {\em non-conservative} torque, providing the possibility of energy exchange
between the conduction electrons and the molecule.

\begin{figure}[t]
\includegraphics[height=3.3cm,keepaspectratio=true]{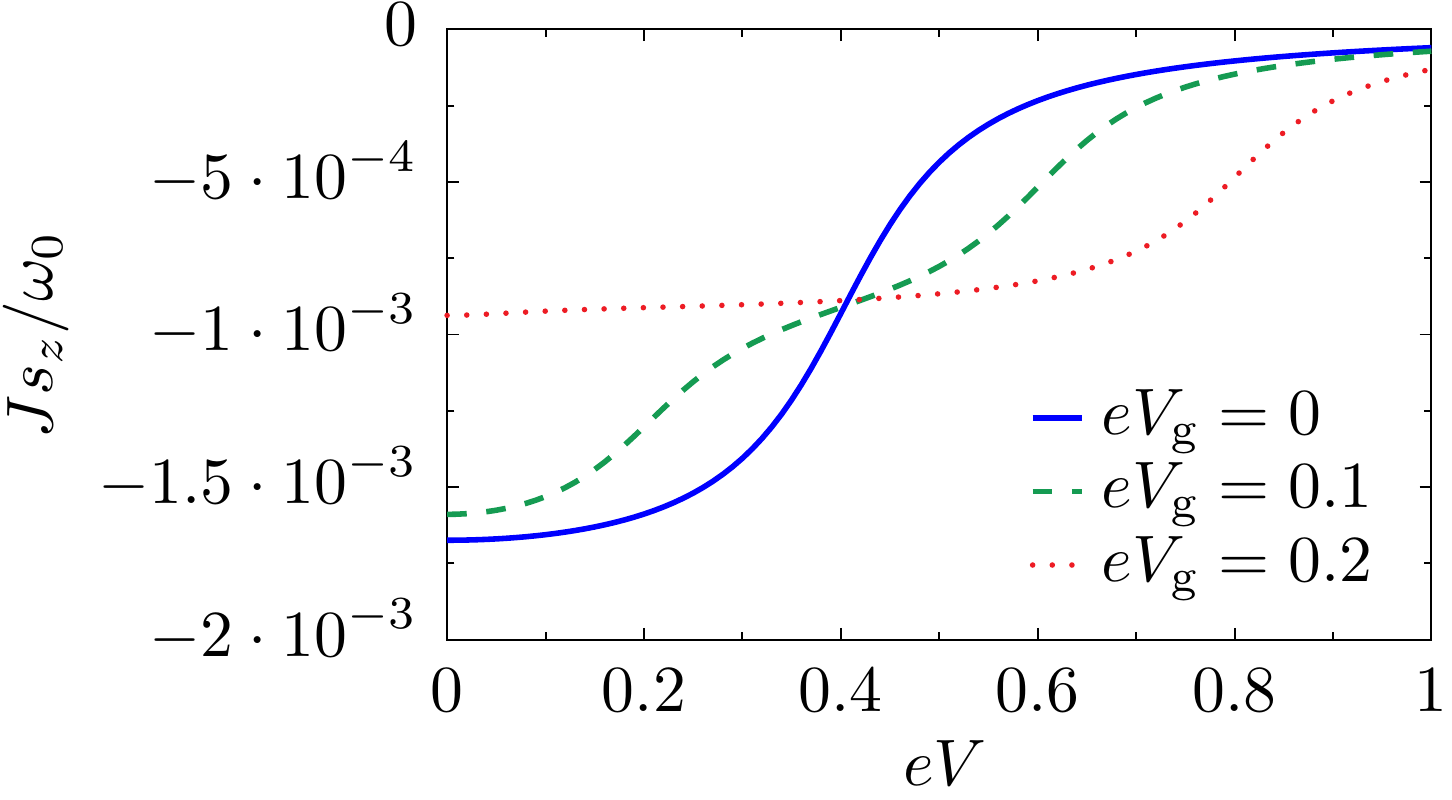}
\caption{(Color online) Component of the average current-induced torque in a uniaxial
situation for unpolarized leads. $J s_z(\mathbf{M})$ is
plotted as a function of the applied bias voltage for different
gate voltages $e V_{\mathrm{g}}$. As discussed in the text, $J
s_z$ changes when the number of
levels in the current window varies at $V_{\mathrm{g}} \pm eV/2 = \pm b$ (see
also Fig.\ \ref{fig:sketch01}). The
plots are obtained at zero temperature at the potential minimum $\vartheta=0$
for the values $J
M/2=0.2$, $\Gamma=0.1$,
$g_e B/2=0.002$ and $g_e=g_d$. The precession
frequency in the absence of coupling to electron spin and magnetic field is $\omega_0=2 DM =0.01$. All energies are measured in
units of the barrier height without magnetic field $D M^2$.}
\label{fig:Jsz}
\end{figure}
\begin{figure*}[t]
\includegraphics[height=3.3cm,keepaspectratio=true]{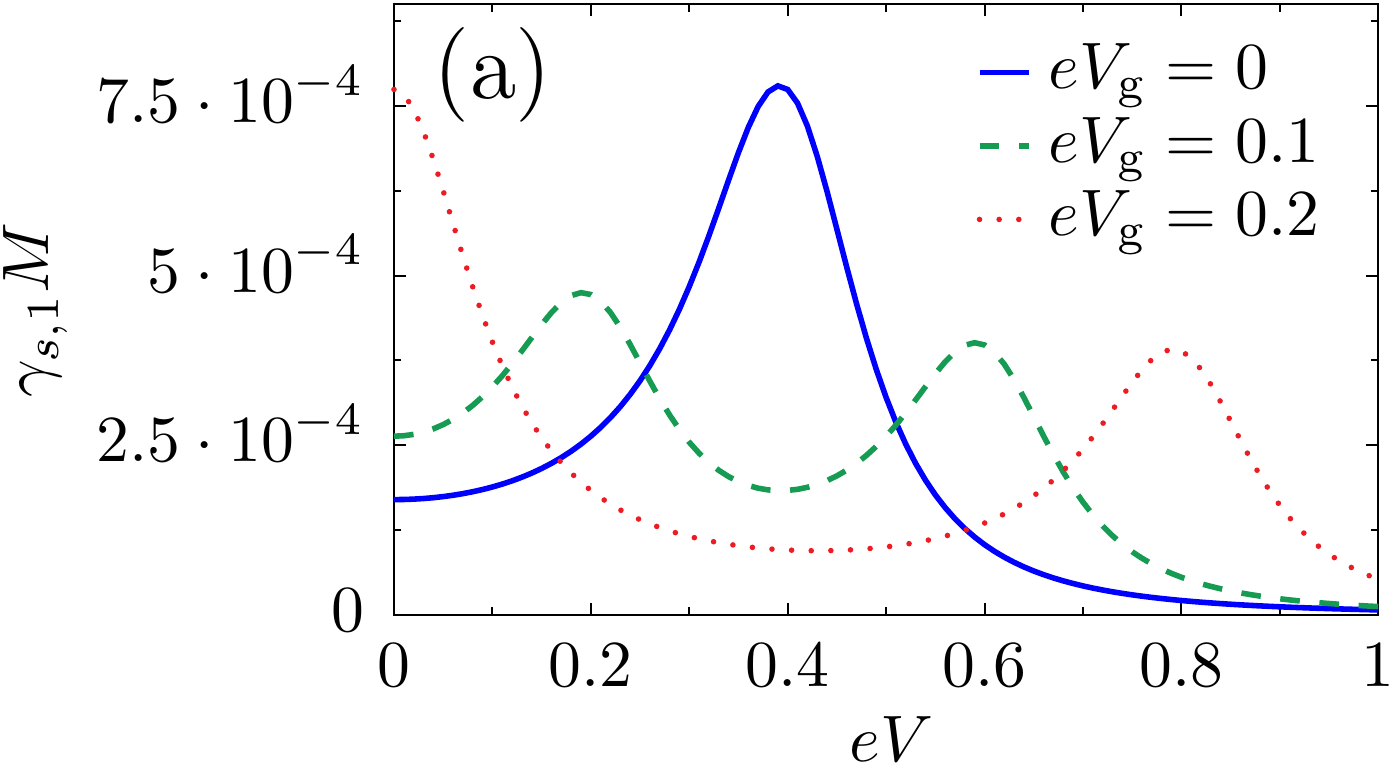}\,\,\,\,
\includegraphics[height=3.3cm,keepaspectratio=true]{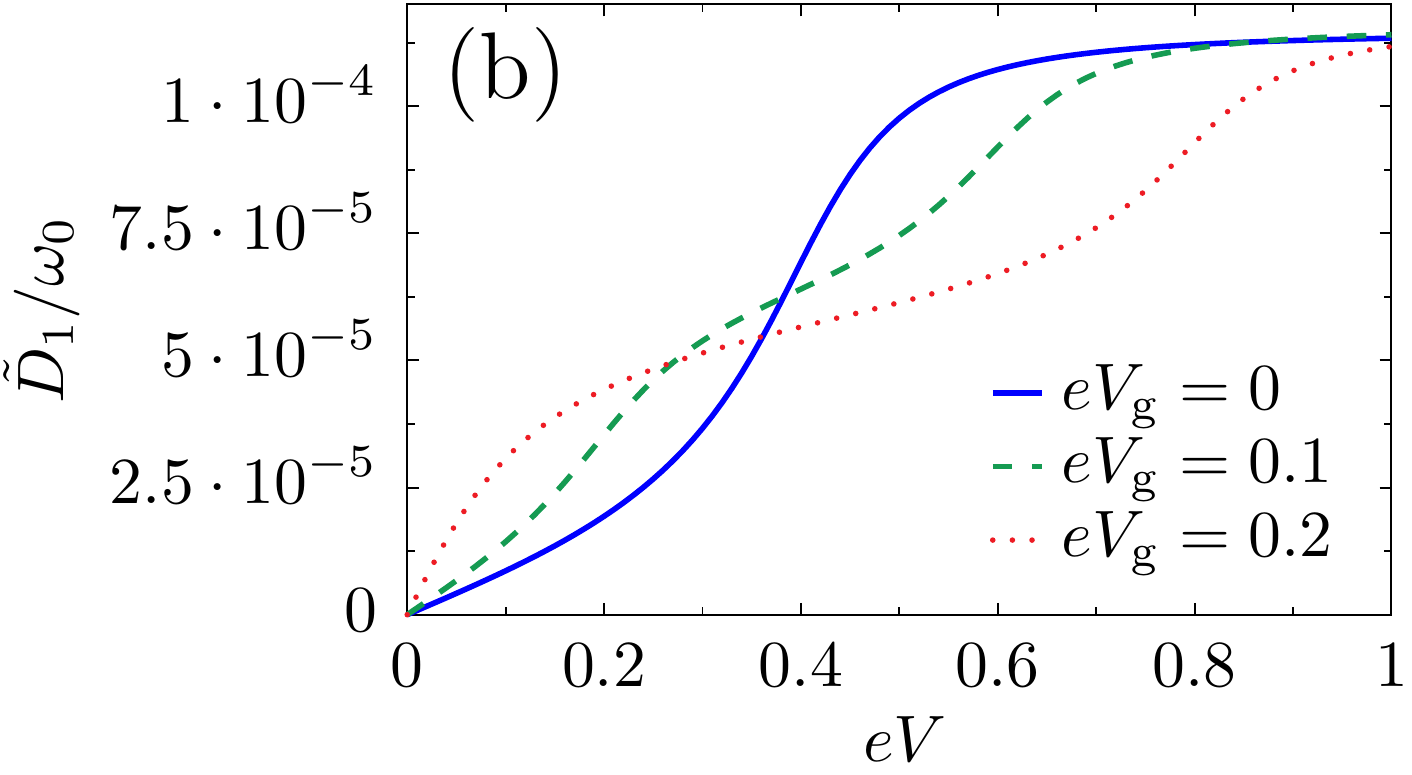}\,\,\,\,
\includegraphics[height=3.3cm,keepaspectratio=true]{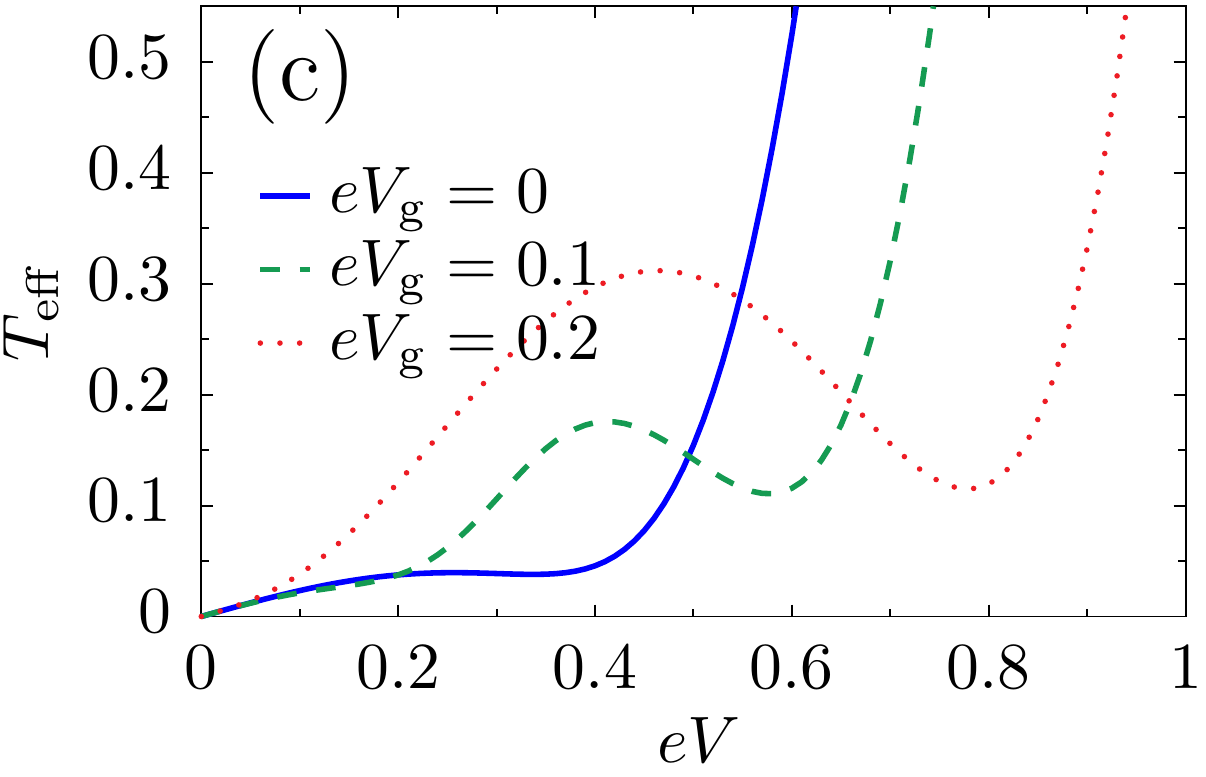}
\caption{(Color online) Damping and fluctuations as a function of the applied
bias voltage. (a) The damping
coefficient $\gamma_{s,1}$ is plotted for three different gate voltages. It is
peaked when the number of
levels in the current region changes
from zero to one to two at $V_{\mathrm{g}} \pm eV/2 = \pm b$ (see also Fig.\ \ref{fig:sketch01}). (b) At these points the fluctuation
kernel $\tilde{D}_1$ increase steplike. The level
broadening results from the
interaction with the leads encapsulated in $\Gamma$. 
(c) The effective
temperature $T_{\mathrm{eff}} = \tilde{D}_1/(2\gamma_{s,1})$ is shown as a
function of the bias voltage. 
The
plots are obtained at the potential minimum $\vartheta=0$
with the
same parameters as in Fig.\ \ref{fig:Jsz}. All energies are measured in units of
the barrier height without magnetic field $D M^2$.} \label{fig:normalleads01}
\end{figure*}
Concrete expressions for these contributions to the current-induced torque can
be obtained from 
\begin{align}
 s_z ({\bf M})  =&  -i \int \frac{d \omega}{2 \pi}  [
G^<_z({\bf M}, \omega) \nonumber \\
&+   G^<_b ({\bf M}, \omega) \Bigl(\frac{g_e B}{2} + \mbox{Re}[\Sigma_s^R] 
\Bigr) ] ,
\label{eq:s_z}\\
s_t ({\bf M}) =&  -\frac{i J}{M} \int \frac{d \omega}{2 \pi} G^<_t
({\bf M}, \omega),\label{eq:s_t}
\end{align}
as derived by substituting Eq.\ \eqref{eq:G0<app} into Eq.\ \eqref{s0} and taking into account possibly
spin-polarized leads with the notation $\Sigma^R_{\alpha,c(s)}=
[\Sigma^R_{\alpha,\uparrow} \pm \Sigma^R_{\alpha,\downarrow}]/2$ for the self-energies.

\begin{figure}[b]
\includegraphics[height=3.3cm,keepaspectratio=true]{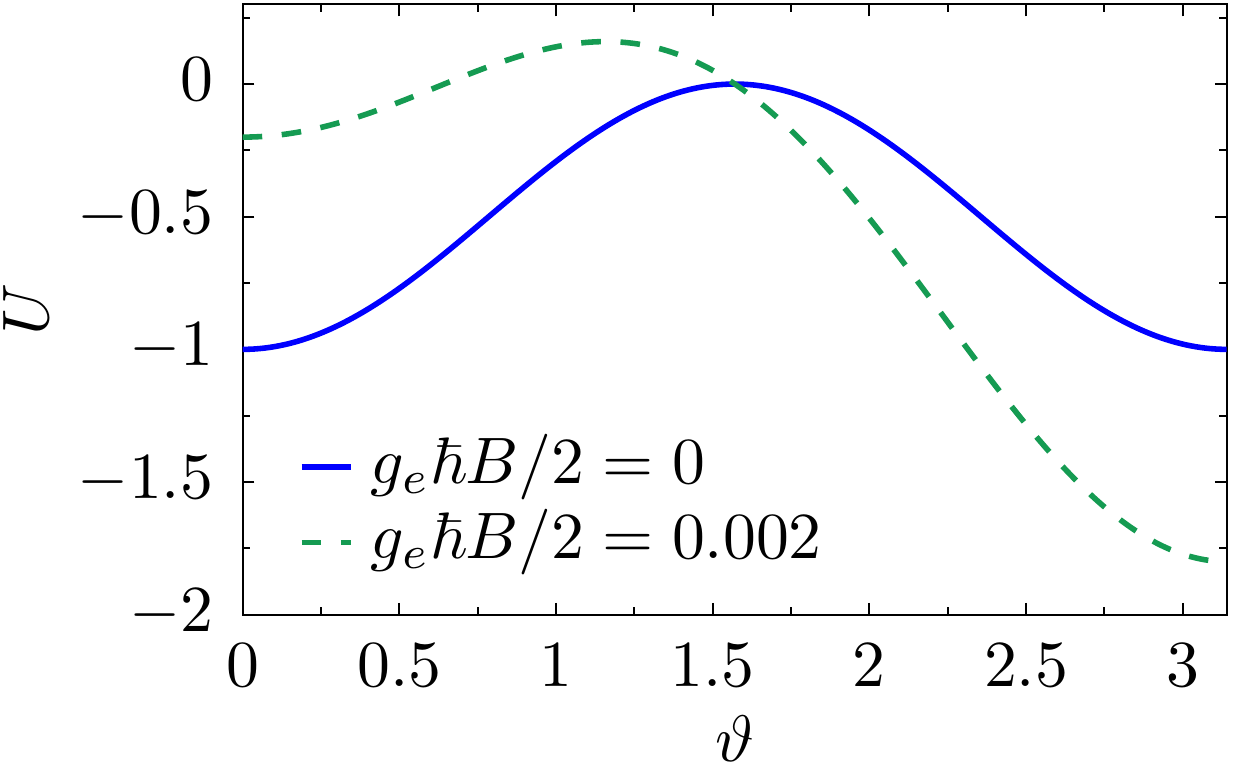}
\caption{(Color online) Potential
experienced by the molecular spin in the absence of coupling to the electrons.
$U(\vartheta)$ is shown for different magnetic fields. All energies
are measured in
units of the barrier height without magnetic field $D M^2$.}\label{fig:U01}
\end{figure}
\begin{figure*}[t]
\includegraphics[height=3.3cm,keepaspectratio=true]{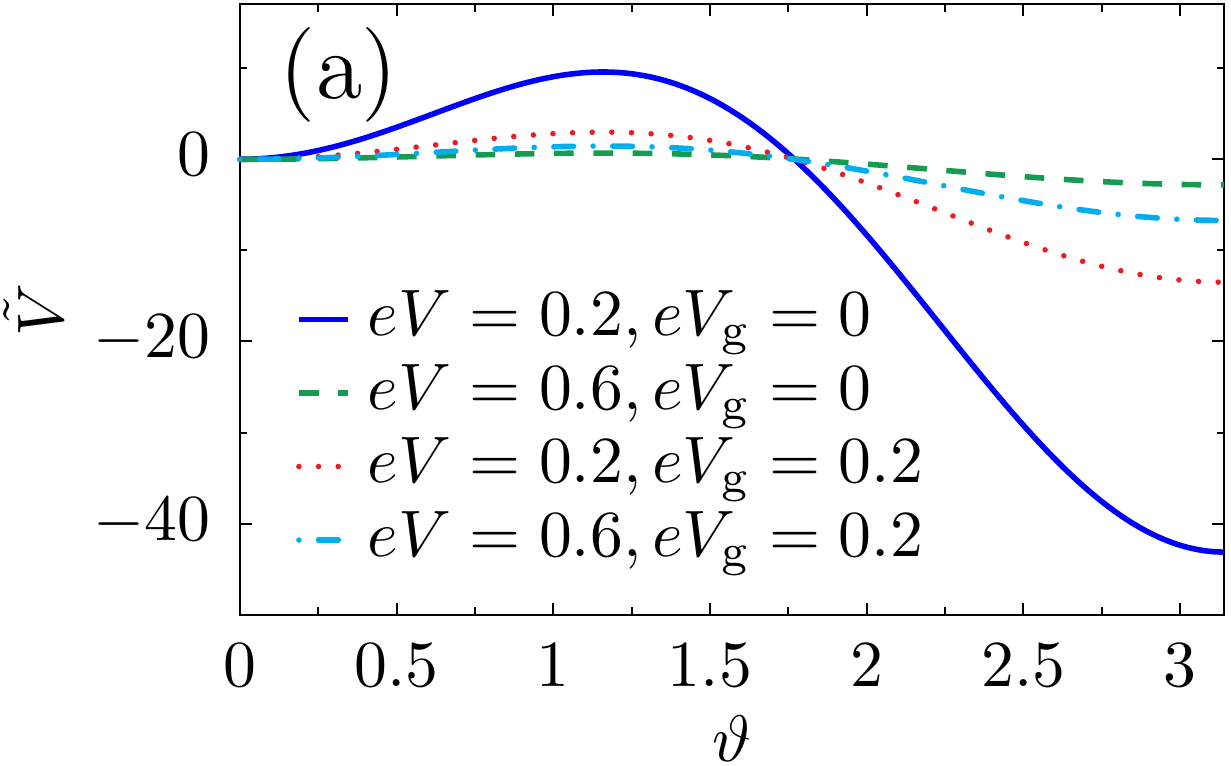}
\includegraphics[height=3.3cm,keepaspectratio=true]{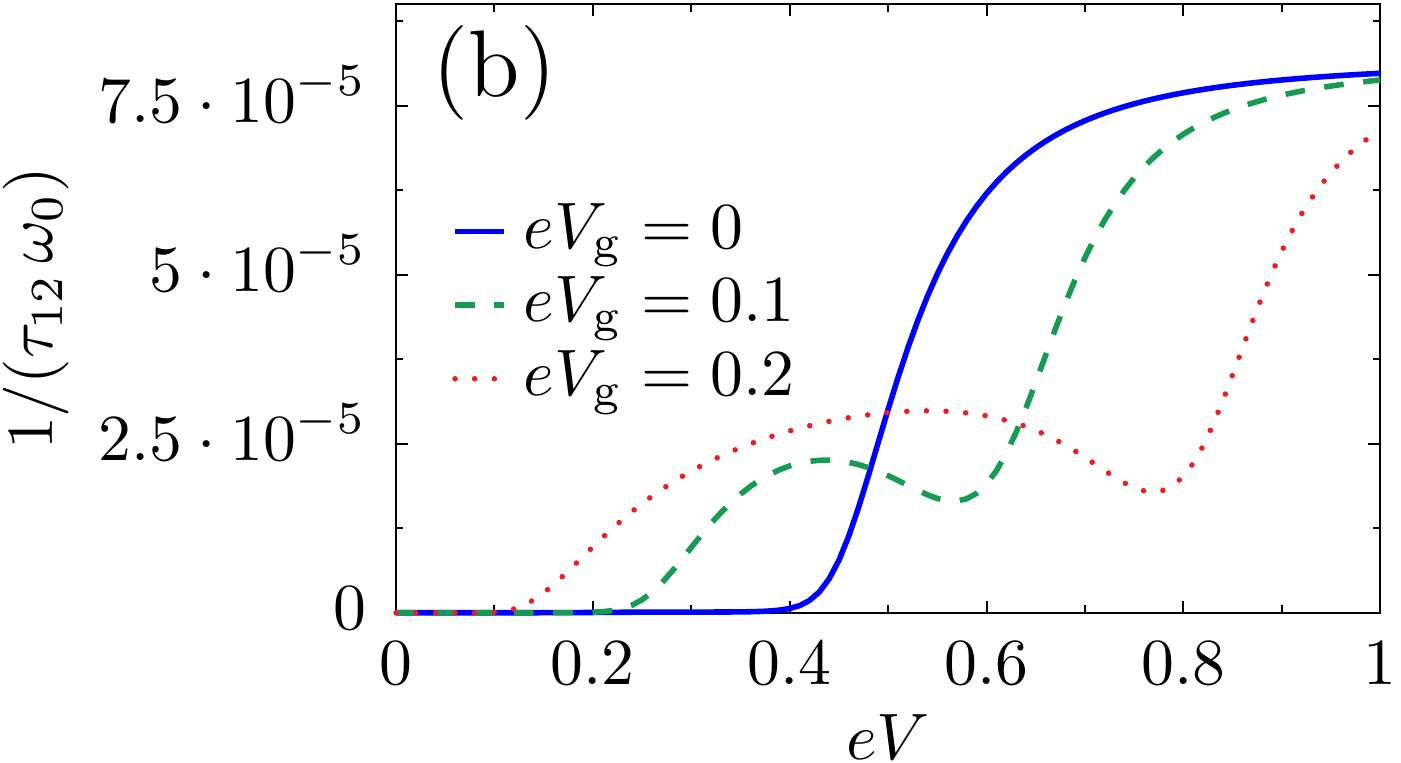}
\includegraphics[height=3.3cm,keepaspectratio=true]{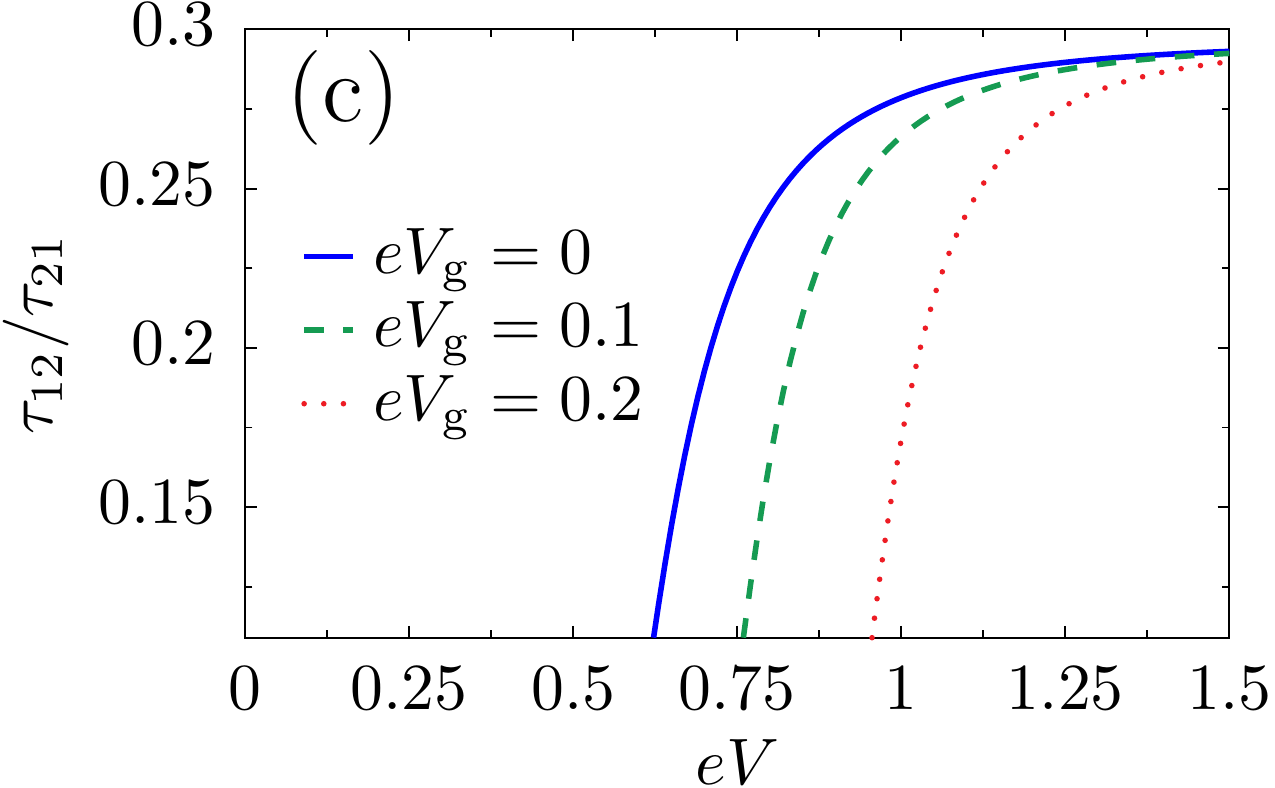}
\caption{(Color online) Switching dynamics for unpolarized leads. (a)
$\tilde{V}$ is plotted for different bias and gate
voltages. (b) The switching rate $1/\tau_{12}$ is plotted as a
function of the applied bias voltage for different
gate voltages. (c) The ratio
between the switching rates $1/\tau_{21}$ and $1/\tau_{12}$ is shown. The
plots are obtained with the same parameters as in Fig.\ \ref{fig:Jsz}. All
energies are measured in
units of the barrier height without magnetic field $D M^2$.}
\label{fig:Vandrates}
\end{figure*}
These general expressions simplify significantly for unpolarized leads,
which corresponds to $\Sigma^R_{\alpha,s}=0$. Indeed, one finds that then
$G^<_t$ and
$G^<_z$ vanish, see Eq.\ \eqref{ges}. This implies in particular that the
component $s_t$ of the average torque vanishes. The remaining conservative
contribution $s_z$ is then found to be $s_z= s_z^- - s_z^+$ with
\begin{align}\label{eq:szT0}
 s_z^\pm(\mathbf{M}) =&\sum_\alpha \frac{\Gamma_\alpha}{\pi
\Gamma} \Bigl( \arctan\left[\frac{\mu_\alpha-\epsilon \pm
b}{\Gamma/2}\right]+\frac{\pi}{2}\Bigr)\frac{g_e B}{2 b}.
\end{align}
Here we assume the limit of zero temperature and introduce the shorthand
$\epsilon= \epsilon_0+\mbox{Re}[\Sigma_c]$.

It is instructive to study the dependence of the average torque on bias and
gate voltage. Notice that,
due to the effective magnetic field acting on the electron spin, the electronic
level splits, see Fig.\ \ref{fig:sketch01}. The average torque is
finite when just one level, corresponding to e.g.
spin-up electrons, is occupied.
In contrast, for sufficiently high bias voltages both spin-up
and
spin-down electrons participate in the transport so
that no net electron spin
acts on the molecule. This is illustrated in Fig.\ \ref{fig:Jsz}, where the average
electronic spin on the molecule is plotted as a function of the applied bias voltage $eV$
for three different values of the molecular level $\epsilon_0$ (as tunable by the gate voltage $e V_{\mathrm{g}}$).

For Gilbert damping and the fluctuating torque, we restrict ourselves to
unpolarized leads. This choice is motivated by the fact that switching of the
molecular spin (as discussed in the next two sections) is dominated by the
average torque for polarized leads (and thus weakly affected by higher orders in
the adiabatic expansion) and by the fluctuating force for unpolarized leads. (We
mention in passing that expressions for Gilbert damping and fluctuating force 
for polarized leads can be readily derived but are rather cumbersome.) 

For unpolarized leads, we can split
the Gilbert damping tensor into one part proportional to the unit matrix and
another proportional to a
projector onto the $z$-axis,
\begin{align}
   {\bf M} \times  \gamma_s \dot{{\bf M} }  =  \gamma_{s,1} {\bf M} \times
\dot{\mathbf{M}}+ \gamma_{s,2} (\dot{\mathbf{M}}\cdot\hat{\mathbf{e}}_{z}) {\bf
M} \times \hat{\mathbf{e}}_z,\label{eq:GilbertN}
\end{align}
where $\gamma_{s,1}$ and $\gamma_{s,2}$ are
scalars. The first term in Eq.\ \eqref{eq:GilbertN} tends
to (anti-)align the molecular spin with the anisotropy axis while
the second modifies the precession frequency.

The coefficients $\gamma_{s,1}$ and $\gamma_{s,2}$ are calculated by inserting
$G_0^<$
and $G_0^>$ from Eq.\ \eqref{ges} into Eq.\ \eqref{eq:gammasN}, resulting in
\begin{align}\label{eq:gammas1}
  &\gamma_{s,1}(\mathbf{M}) = \int \frac{d \omega}{8 \pi}
\sum_{\alpha \beta}  \frac{J^2\,\Gamma_\alpha \Gamma_\beta (-\partial_\omega
f_\beta)}{\prod_{\pm}\left[(\omega-\epsilon \pm b)^2+(\Gamma/2)^2\right]},
\end{align}
and
\begin{align}\label{eq:gammas2}
  &\gamma_{s,2}(\mathbf{M}) = \int \frac{d
\omega}{16 \pi}
\sum_{\alpha \beta}  \frac{(g_e B J)^2\,\Gamma_\alpha \Gamma_\beta
(-\partial_\omega
f_\beta)(\omega-\epsilon)^2}{\prod_{\pm}\left[(\omega-\epsilon \pm
b)^2+(\Gamma/2)^2\right]}.
\end{align}
The damping coefficient is peaked when the number of
levels between $\mu_L$ and $\mu_R$ changes and thus vanishes at large
voltages when both levels are in the transport window. We illustrate this
dependence of $\gamma_{s,1}$ on gate and bias voltage in
Fig.~\ref{fig:normalleads01}.
The prefactor $\alpha$ in Eq.\ \eqref{eq:LLG2} is calculated in the same way as the damping coefficients, and
the resulting expression is relegated to the appendix, see Eq.\ \eqref{eq:CN}.

We close this section with the corresponding expression for the variance of the
fluctuating Zeeman field, Eq.\ \eqref{eq:variance}, which becomes
$\tilde{D}^{kl}(\mathbf{M})=\tilde{D}_1(\mathbf{M}) \delta_{kl}  +
\tilde{D}_2(\mathbf{M}) b_k b_l$, where
\begin{align}\label{eq:DN}
  \tilde{D}_1(\mathbf{M}) &= \frac{J^2}{2} \int \frac{d \omega}{2 \pi}
\sum_{\alpha \beta}
\frac{\Gamma_\alpha \Gamma_\beta f_\alpha (1-
f_\beta)}{\prod_{\pm}\left[(\omega-\epsilon \pm
b)^2+(\Gamma/2)^2\right]}\nonumber\\
    \tilde{D}_2(\mathbf{M}) &= J^2 \int \frac{d \omega}{2 \pi} \sum_{\alpha
\beta}  
\frac{\Gamma_\alpha \Gamma_\beta f_\alpha (1-
f_\beta)\,(\omega-\epsilon)^2}{\prod_{\pm}\left[
(\omega-\epsilon \pm b)^2+(\Gamma/2)^2\right]^2},
\end{align}
for unpolarized leads.
As illustrated in Fig.\ \ref{fig:normalleads01}, the strength of the
fluctuations
changes with the number of electronic levels in the transport window and
saturates at high bias voltages when both levels lie
within.

\section{Fluctuation-induced switching of the molecular moment for unpolarized
leads.}
\label{sec:Fluctuation}
We now apply our results to discuss the switching dynamics for unpolarized
leads. In the absence of coupling to the electrons the
molecular
spin moves in the potential $U=  g_d B M_z -D M_z^2$. For sufficiently small
magnetic fields, two minima are present, corresponding to parallel and
antiparallel alignment of the spin to the magnetic
field, see Fig. \ref{fig:U01}.

Assume that the molecular spin is initially aligned parallel to the magnetic
field. Due to the interaction with the electrons the molecular spin fluctuates
about this initial state, causing spin flips at a certain rate which we
calculate in this section. Clearly, these fluctuations depend on
temperature and applied bias voltage. If the system is in thermal equilibrium,
this
is a standard problem.\cite{brown-1963} Our approach allows us to
extend these standard results to out-of-equilibrium situations in the presence
of a bias voltage in addition to finite temperature.  We also demonstrate that
the
orientation of
the molecular spin can be read out by tracking the current
through the molecule.

\begin{figure*}[tb]
\includegraphics[height=3.3cm,keepaspectratio=true]{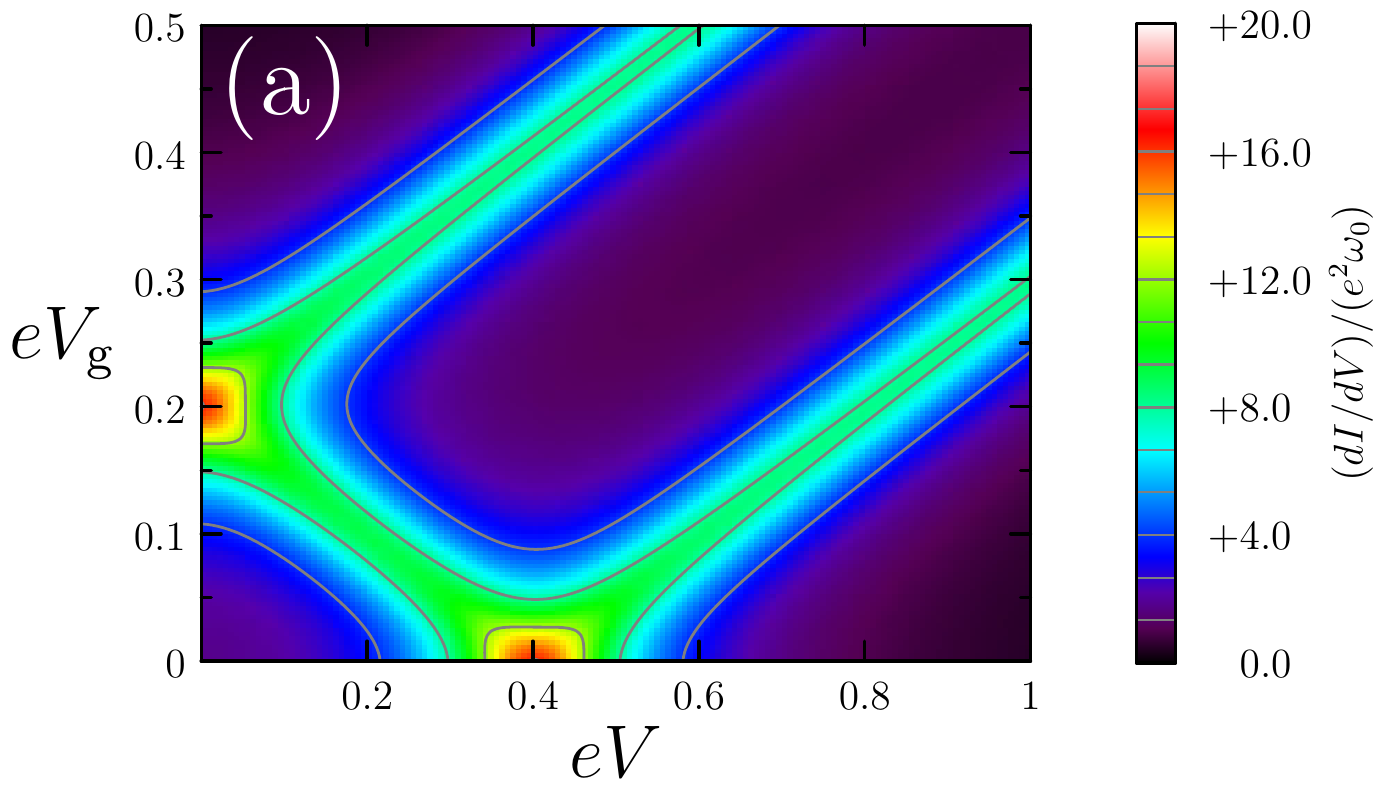}\,\,\,\,
\includegraphics[height=3.3cm,keepaspectratio=true]{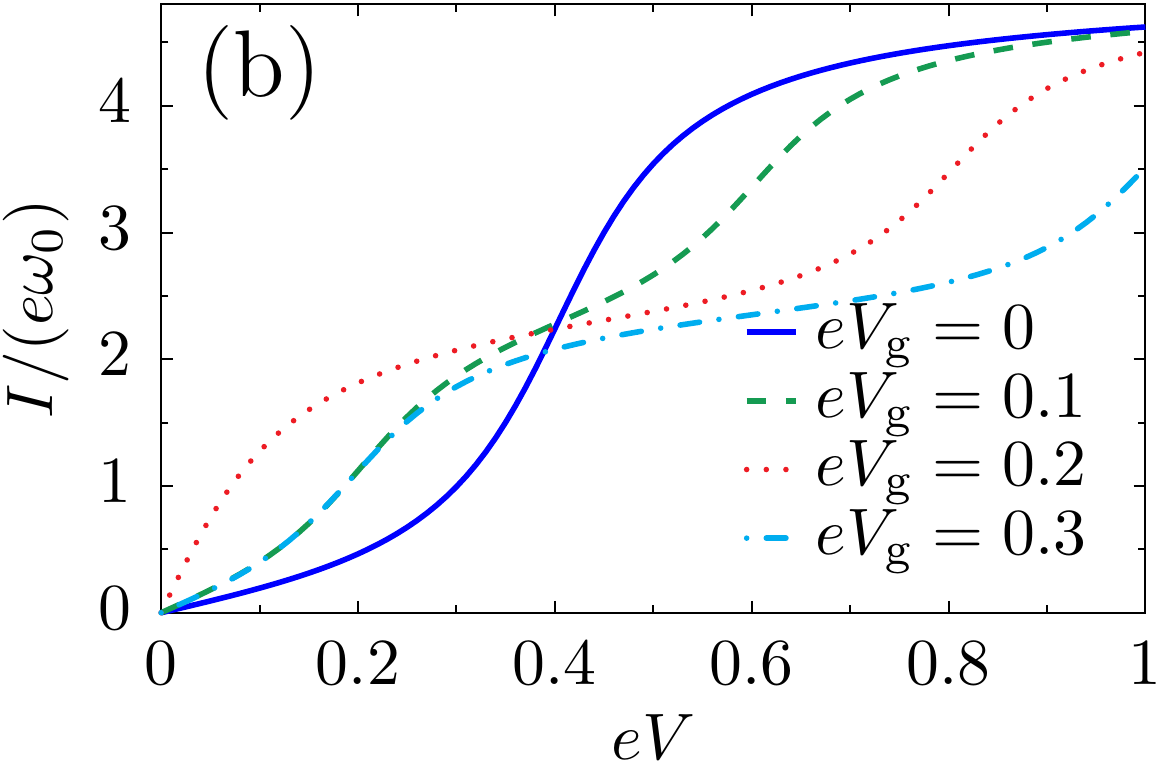}\,\,\,\,
\includegraphics[height=3.3cm,keepaspectratio=true]{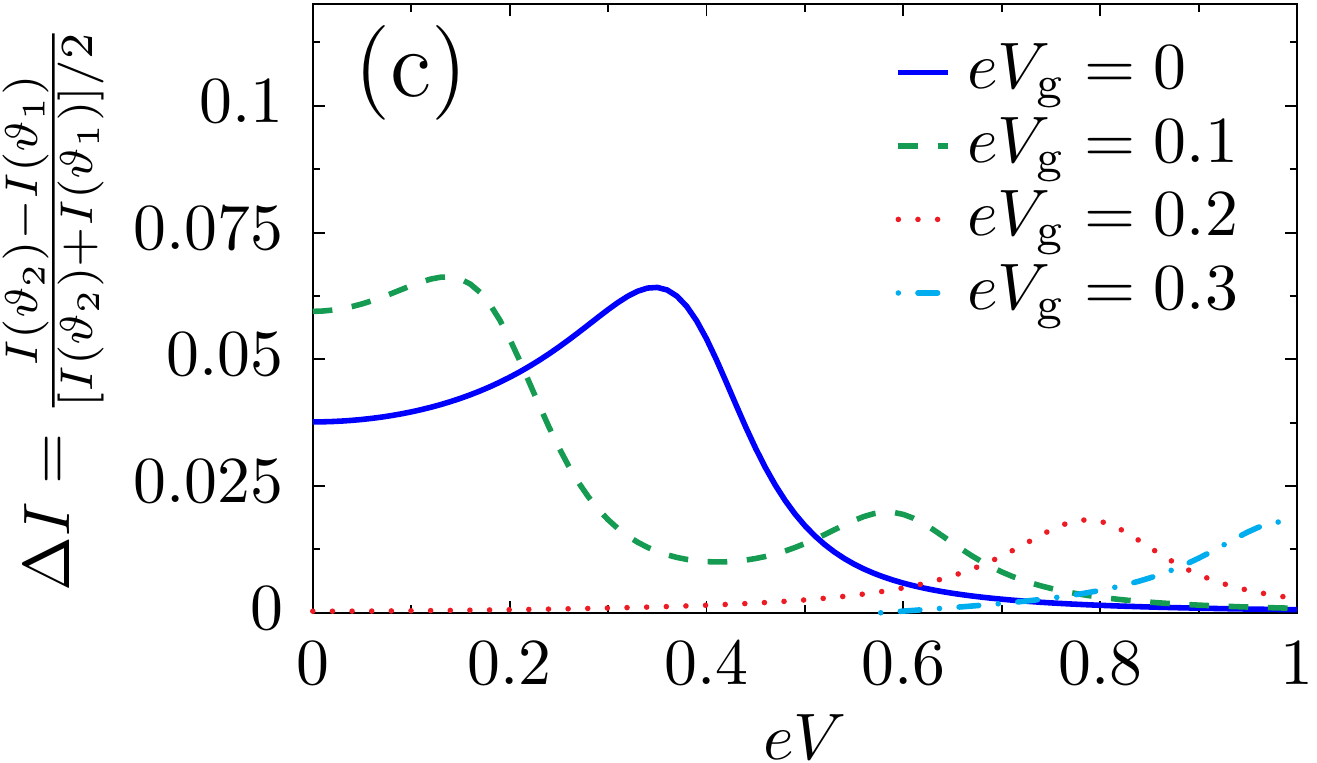}
\caption{(Color online) Current-voltage characteristics. (a) The differential conductance $dI/dV$ is peaked when the number
of levels in the current window changes at $V_{\mathrm{g}} \pm eV/2 = \pm b$.
(b) Obviously, the current changes when the number of level with energy between
the chemical potentials of the two leads changes. (c) The level splitting $2
b(\vartheta)$ depends on the orientation of the molecular spin. The
relative difference of the current evaluated at
the two potential minima is plotted as a function of the bias voltage. [(a) and (b) are
evaluated at $\vartheta=0$. The
plots are obtained with the same parameters as in Fig.\ \ref{fig:Jsz} and all energies are measured in units of the barrier height without magnetic field $D M^2$.]}
\label{fig:current}
\end{figure*}
\subsection{Fokker-Planck equation}
Our approach is based on an equivalent Fokker-Planck formulation of the Langevin
dynamics of the molecular spin. We first rewrite the Langevin equation
\eqref{eq:LLG2} for unpolarized leads. Describing the orientation of the
molecular spin in terms of a polar angle $\vartheta$, measured relative to the
applied magnetic field, and an azimuthal angle $\varphi$, and noting that
$\dot{\mathbf{M}}/M=\dot{\vartheta}\hat{
\mathbf{e}}_{\vartheta}+\dot{\varphi}
\sin\vartheta\hat{\mathbf{e}}_{\varphi}$, we find the Langevin equation
\begin{align}
  \dot{\vartheta} &= \alpha \left[M \gamma_{s,1} \sin \vartheta \,
\dot{\varphi} - \delta {\cal
B}_\varphi\right] \nonumber\\
   \sin\vartheta \, \dot{\varphi} &= \alpha \bigl[ - \partial_\vartheta U/M - J
s_z \sin\vartheta+
\delta {\cal
B}_\vartheta \nonumber\\ 
  &\qquad - M
\left(\gamma_{s,1} + \gamma_{s,2} \sin^2 \vartheta \right)
\dot{\vartheta}\bigr], \label{eq:LLGpolar1}
\end{align}
where the noise correlator is given in polar
coordinates by $\tilde{D}^{\varphi \varphi} = \tilde{D}^{\vartheta \varphi} =
\tilde{D}_1$ and $\tilde{D}^{\vartheta \vartheta} = \tilde{D}_1 +  \left[(g_e
B/2) \cos\vartheta \sin\vartheta\right]^2 \tilde{D}_2$, with $\tilde{D}_i$
defined in Eq.\ \eqref{eq:DN}.

Following standard procedures,\cite{zwanzig-2001} we now derive the
corresponding Fokker-Planck equation  for the probability distribution
$P(\mathbf{M},t)$ of the magnetization vector $\mathbf{M}$ at time $t$. In the
uniaxial situation under consideration, this probability distribution is
independent of $\varphi$ and depends on the angle $\vartheta$ only. As outlined
in Appendix \ref{app:fp-derivation} for the convenience of the reader, we then
obtain the Fokker-Planck equation
\begin{align}
  \partial_t P(\vartheta,t) &= \frac{1}{\sin\vartheta} \partial_\vartheta \sin\vartheta
e^{-\tilde{V}(\vartheta)}\partial_\vartheta
e^{\tilde{V}(\vartheta)}\tilde{\beta}(\vartheta)\,
P(\vartheta,t) \label{eq:FPtheta}.
\end{align}
This equation has the stationary solution
$P(\vartheta)_{\mathrm{stat}} \propto
\exp[{-\tilde{V}(\vartheta)}]/\tilde{\beta}$. 
Here we have introduced
\begin{align}
  \tilde{\beta}(\vartheta) &=
\frac{\alpha^2 \tilde{D}_1/2}{1+\alpha^2M^2
\gamma_{s,1}\left(\gamma_{s,1} + \gamma_{s,2} \sin^2\vartheta\right)},
\end{align} 
and
\begin{align}
    \tilde{V}(\vartheta)&=\int^\vartheta d\vartheta'\,
\frac{\partial_{\vartheta'} U  + M  s_z J
\sin\vartheta'}{\tilde{D}_1/(2\gamma_{s,1})}.
\end{align}

As long as the anisotropy is sufficiently large, $U(\vartheta)$ has a minimum
$U_1$ at $\vartheta=\vartheta_1=0$, another minimum $U_2$ at 
$\vartheta=\vartheta_2=\pi$ and a maximum
$U_m$ at $0<\vartheta<\pi$. We assume that this holds also for
$\tilde{V}(\vartheta)$ and visualize the dependence of
$\tilde{V}(\vartheta)$ on gate and bias voltage in Fig.\
\ref{fig:Vandrates}. One clearly sees that the difference between the values of
$\tilde{V}$ at the minima and the maximum decreases with increasing bias
voltage, as one expects from the behavior of fluctuations and damping, cf.
Fig.\ \ref{fig:normalleads01}.

Note that in equilibrium the ratio $\tilde{D}_1/(2\gamma_{s,1})=T$, as dictated by
the fluctuation-dissipation
theorem. 
For zero temperature but finite bias voltages $V$ it is
sometimes instructive to interpret this ratio as an effective temperature in
each potential well,
$T_{\mathrm{eff}}\simeq\tilde{D}_1/(2\gamma_{s,1})$ (as done for instance in
Refs.\ \onlinecite{mozyrsky-2006,pistolesi-2008,nunez-2008}), see Fig.\ \ref{fig:normalleads01}.
Generally however, both
coefficients, 
$\tilde{D}_1$ and $\gamma_{s,1}$ are angle dependent and non trivial
functions of voltage, as we have seen explicitly above.

We calculate how long the molecular spin remains on one half of the Bloch
sphere. The
mean time between passing the energy barrier $\tau_{ij}$ due to the interaction
with the
electrons is then found by a standard procedure.\cite{zwanzig-2001} We consider
an
adjoint
equation to Eq.\ \eqref{eq:FPtheta}, 
\begin{align}
\frac{\tilde{\beta}(\vartheta)}{\sin\vartheta}e^{\tilde{V}(\vartheta)}
\partial_\vartheta e^{-\tilde{V}(\vartheta)}\sin\vartheta \partial_\vartheta
\,\tau_{ij}(\vartheta) = -1/2,
\end{align}
with an absorbing boundary condition $\tau_{ij}(\vartheta_m)=0$ in order to get
the mean first passage time, as briefly outlined in Appendix \ref{app:fp-derivation}. The factor
$1/2$ takes into account that it is
equally likely to go to $\vartheta \gtrless \vartheta_{\mathrm{m}}$ at
$\vartheta = \vartheta_{\mathrm{m}}$.
Solving the equation yields
\begin{align}
  \tau_{12}(\vartheta) = 2
\int_{\vartheta}^{\vartheta_m}d\vartheta'\,
\frac{e^{\tilde{V}(\vartheta')}}{\sin\vartheta'}\int_{\vartheta_{1}}^{\vartheta'
} d\vartheta''\,\frac{\sin\vartheta''}{\tilde{\beta}(\vartheta'')}
e^{-\tilde{V}(\vartheta'')}\label{eq:tau12}
\end{align}
for passing from $\vartheta <
\vartheta_m$ to $\vartheta > \vartheta_m$ and an analogous expression
for the opposite process. 

When the potential minima are well separated and the
fluctuations are small, we
can give an analytical expression for the switching rate. In this limit, the
integrals in
\eqref{eq:tau12} can be evaluated by saddle-point integration (cp. Ref.
\onlinecite{brown-1963} for the situation in which the coefficients do
not depend on $\vartheta$), yielding
\begin{align}
  \frac{1}{\tau_{ij}} \simeq \frac{1}{\sqrt{2\pi}}  \sin\vartheta_m
\sqrt{|\tilde{V}''(\vartheta_m)|}\tilde{\beta}(\vartheta_i)
\tilde{V}''(\vartheta_i)
e^{-[\tilde{V}(\vartheta_m)-\tilde{V}(\vartheta_i)]}.
\end{align}
Hence, the rate depends exponentially on the difference between
$\tilde{V}$
evaluated at its maximum and minimum, respectively, so that it can be tuned by
varying bias voltage and gate potential. The general behavior of $1/\tau_{ij}$,
as given by Eq.\ \eqref{eq:tau12},
is shown in Fig.\ \ref{fig:Vandrates} for typical values  as a function of gate
and bias voltages. We discussed above that the fluctuations increase with the
number of levels in the current window. This is also reflected in the
fluctuation induced transition rates which increase with the
bias voltage accordingly.

\subsection{Current}
The current through lead $\alpha$ is given by the change of the number of
particles in the lead times the electronic charge, $I_\alpha = - i
e \langle \left[H,\sum_{k_{\alpha},\sigma}
c^{\dagger}_{k_{\alpha},\sigma}c_{k_{\alpha},\sigma}\right]\rangle$. In the
adiabatic limit this becomes\cite{jauho}
\begin{align}
  I_\alpha = e \int \frac{d \omega}{2\pi}
\mathrm{tr}\left[\left(G_0^R - G_0^A\right) \Sigma_\alpha^< + G_0^<
\left(\Sigma_\alpha^A - \Sigma_\alpha^R\right)\right],
\end{align}
where $\Sigma_{\alpha}=\Sigma_{\alpha,c}$ for unpolarized leads.
Noting that $I_L = -I_R = I/2$ and assuming symmetric coupling to the
leads, $\Gamma_L = \Gamma_R = \Gamma/2$, we obtain, by inserting the
expressions for the Green's functions [Eqs.\ \eqref{eq:G0Rapp}
and \eqref{eq:G0<app}] and the self-energies [Eqs.\
\eqref{eq:sigmarapp} and \eqref{eq:sigmaless}] after
straightforward algebra
\begin{align}
  I = \frac{e}{4 \pi} \Gamma \sum_\pm \left[\arctan\left(\frac{\mu_L - \epsilon
\mp b}{\Gamma/2}\right) - (L \leftrightarrow R) \right],
\end{align}
which is valid at zero temperature.
As discussed above, the electronic level splits due to the interaction with the
effective magnetic
field $\mathbf{b}$, defined in Eq.\ \eqref{eq:fieldb1}. When this level splitting is larger than the level broadening $\Gamma$, the current increases
as the number of levels in the transport window increases, see Fig.\
\ref{fig:sketch01}. This is reflected in peaks of the
differential conductance $dI/dV$ as a function of gate and bias voltage. Note
that the splitting of the electronic levels and thus the number of levels in
the transport window depends on the molecular spin orientation since
$b=b(\vartheta)$.
As a consequence, the current is also a function of $\vartheta$. In principle,
this allows one to read out the
molecular switch via current measurements, see Fig.\ \ref{fig:current}.

\begin{figure}[t]
\includegraphics[width=7cm,keepaspectratio=true]{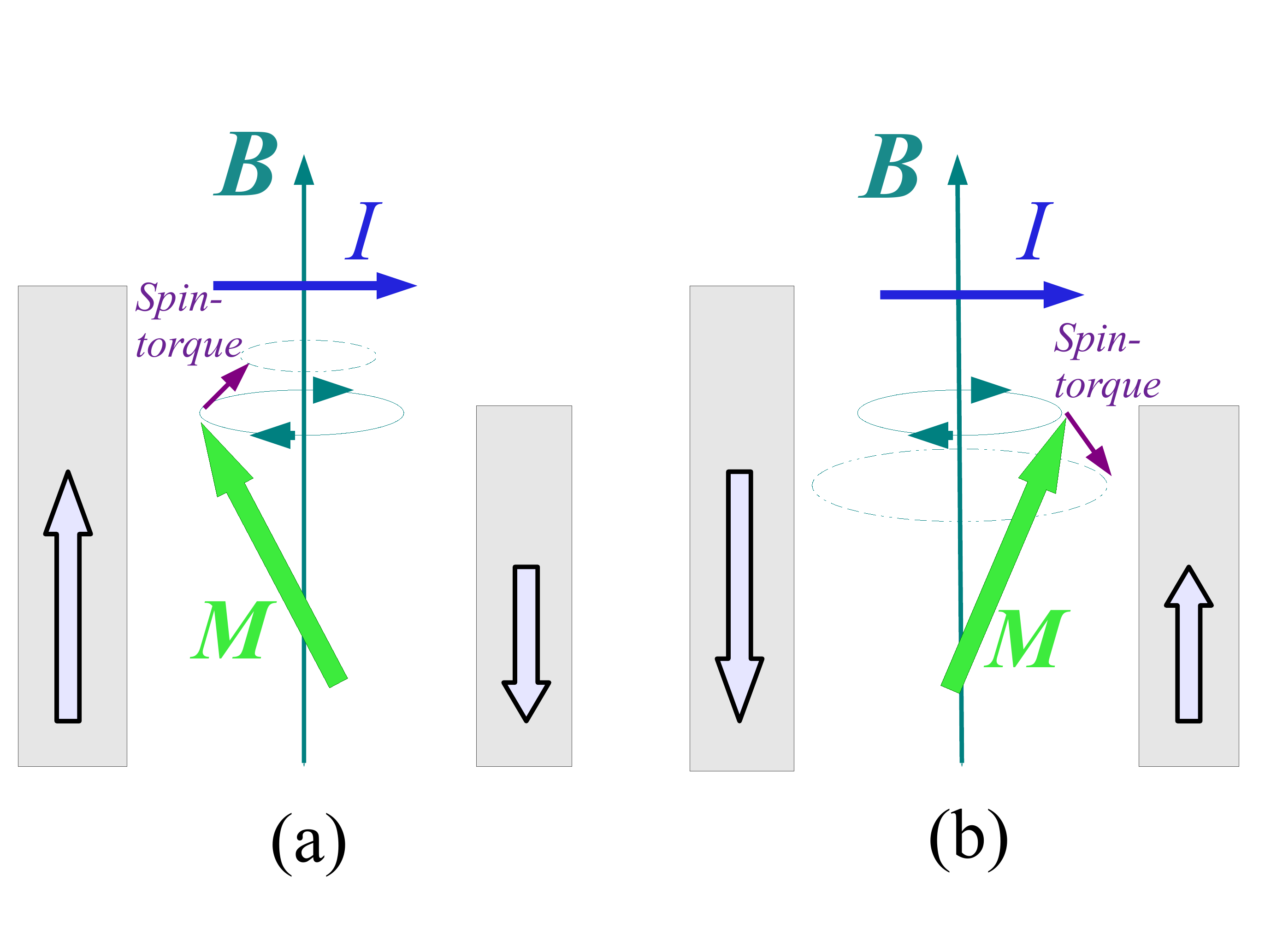}
\caption{(Color online) Sketch of the effect of polarized leads inducing
spin-torque-transfer (indicated by the
magenta arrow)
on the molecular moment. Depending on the polarization and the current, the
spin-torque tends to align the molecular moment either
(a) along or (b) against the magnetic field.} 
\label{fig:sketch}
\end{figure}
\section{Spin-torque-induced switching with polarized leads}\label{sec:Torque}
The switching mechanism discussed in the previous section originates
in fluctuations of the molecular magnetic moment, introduced by the coupling to
the itinerant electrons. In Sec.\ \ref{sec:smat} we have seen that the
presence of
polarized leads opens the possibility of negative Gilbert damping  which could
favor
the
switching of the molecular spin. This mechanism strongly depends on the
details
of the system, like the value of the mean chemical potential $\mu$ and the
applied bias voltage. However, for spin-polarized leads,
switching of the molecular
moment under general non-equilibrium conditions will typically be dominated by a
different mechanism which is driven by the non-conservative (or spin-transfer) torque
exerted by the coupling to the current carrying electrons. The generic effect of
the spin-torque in the dynamics of ${\bf M}$ has been reviewed in
Ref.\ \onlinecite{spin-torque}. This term appears
already in leading order
of the Born-Oppenheimer approximation in which Gilbert damping and fluctuations
can be neglected.

In this section we focus on this spin-torque $-J s_t(\hat{\bf e}_z \times 
\hat{{\bf M}} )$, see Eq.\ \eqref{eq:avtorque}, in the Landau-Lifshitz-Gilbert
equation \eqref{eq:LLG2}, where $s_t$ is given by Eq.\ \eqref{eq:s_t}. We
analyze under which microscopic conditions
it is expected to drive switching in our molecular setup.
In the
present case it is clear that it moves the vector ${\bf M}$ 
along the azimuthal direction, tending to align it along the magnetic field.
Thus, given a tilted molecular
magnetic moment  ${\bf M}$ precessing around the magnetic field, for $s_t <0$
the
spin torque induces a spiral trajectory moving ${\bf M}$ toward orbits of
smaller radius around the magnetic field. Instead, for $s_t >0$ it induces
orbits of larger radius enabling the switching to the opposite
hemisphere, with ${\bf M}$ tending to align opposite to the external
magnetic field.

\begin{figure}[t]
\includegraphics[width=5cm,keepaspectratio=true]{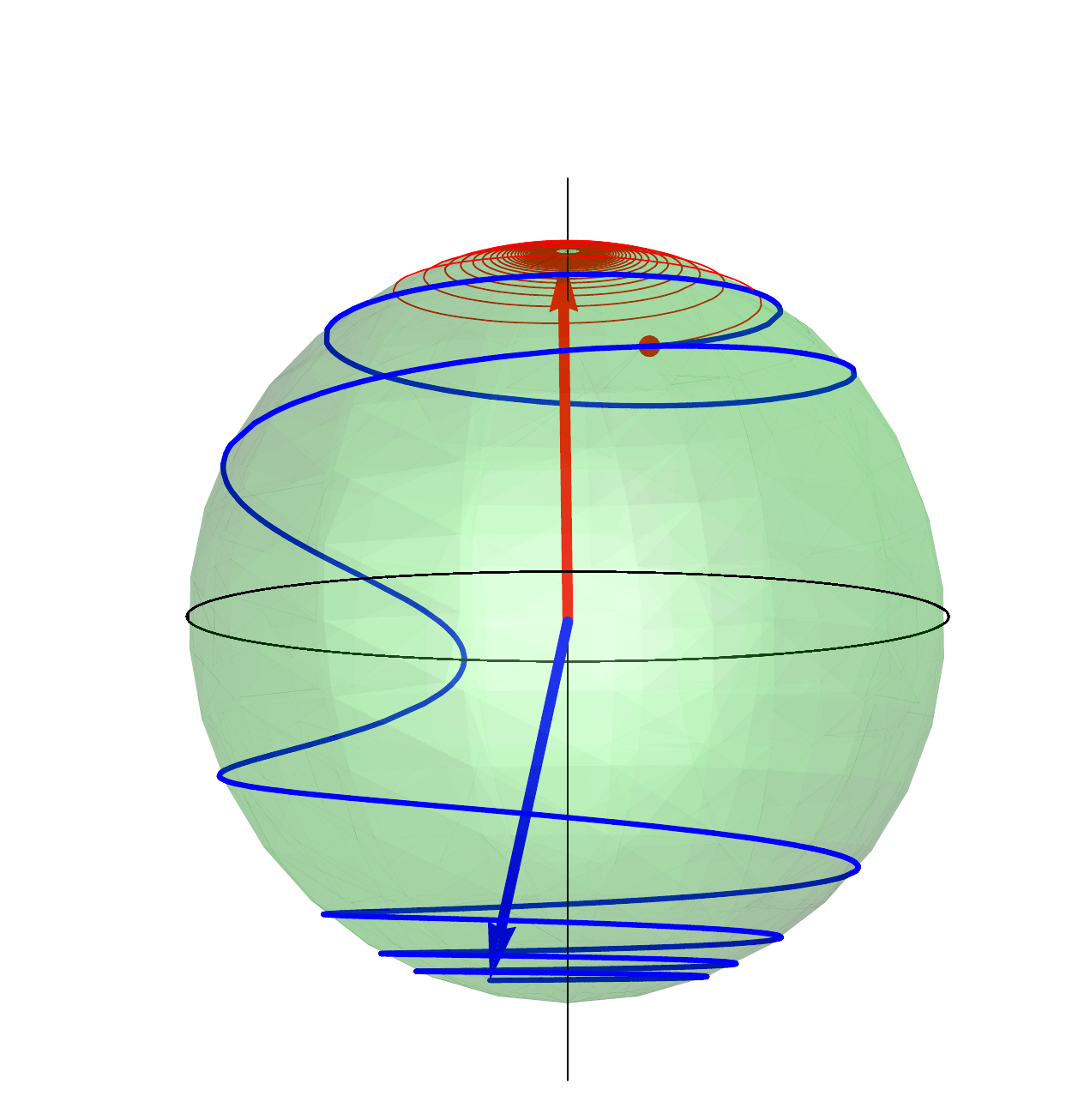}
\caption{(Color online) Motion of the molecular moment in the
presence of spin-polarized leads. For negative bias
voltage $eV =-0.4$, the magnetic moment is driven toward the positive $z$ axis
(red curve) while inverting the voltage $eV\rightarrow - eV$ causes a flip of
the magnetic moment (blue curve). (We consider $\Gamma_L^\uparrow =
\Gamma_R^\downarrow =0.1 $, $\Gamma_L^\downarrow=\Gamma_R^\uparrow=0$ and $e V_{\mathrm{g}}
=0$; the other parameters
are the same as in Fig.\ \ref{fig:Jsz}.)} 
\label{fig:trajectory}
\end{figure}

In our model, the behavior of $s_t$ can be rather easily analyzed in the limit
of completely polarized leads, e.g.
$\Gamma_L^{\uparrow}=\Gamma_R^{\downarrow}=\Gamma/2$. In this limit Eq.\
\eqref{eq:s_t} simplifies to
\begin{align}
  s_t = -\frac{J \Gamma^2}{4\pi M} \int d\omega
\frac{f_L-f_R}{\prod_{\pm}\left[(\omega-\epsilon \pm
b)^2+(\Gamma/2)^2\right]}.
 \end{align}
More generally, the sign of $s_t$ is determined by the condition
 \begin{equation} 
 \mbox{sgn}[s_t] = \mbox{sgn}\left[\left(\Gamma_L^{\downarrow}
\Gamma_R^{\uparrow}-
 \Gamma_L^{\uparrow} \Gamma_R^{\downarrow}\right) \left(f_L-f_R
\right) \right].
 \end{equation}
Thus, when we consider a $\uparrow$-polarized left lead with
$\Gamma_L^{\uparrow} \Gamma_R^{\downarrow} > \Gamma_L^{\downarrow}
\Gamma_R^{\uparrow}$, a current flowing from left
to right, $\mu_L>\mu_R$, results in $s_t<0$ and thus antialignment of magnetic
moment and magnetic field. For the opposite spin-polarization, the spin-torque
tends to align the magnetic moment with the magnetic field, as sketched in Fig.\ \ref{fig:sketch}. 

For a given spin-polarization, inverting the
direction of the current can switch the orientation of the magnetic moment in the same way. This is studied by solving numerically the equation of motion for the molecular spin in the strictly adiabatic limit, hence neglecting Gilbert damping and fluctuations, in the presence of strongly polarized leads. In Fig.\ \ref{fig:trajectory} we show the time evolution of the molecular spin initially slightly deviating from the magnetic field axis for two different voltages. Clearly, the motion of the molecular spin is determined by the direction of the current through the molecule, showing that inverting the bias voltage causes spin-flips in this setup.

\section{Summary and conclusions}
\label{sec:Conclusion}
In this work we have considered an anisotropic magnetic molecule in a
single-molecule junction in which conduction electrons couple via exchange to
the localized magnetic moment. The resulting current-induced torques have been
analyzed by means of the non-equilibrium Born-Oppenheimer approximation,
which gives rise to Langevin dynamics
of the magnetic moment, described by a generalized Landau-Lifshitz-Gilbert
equation. This approximation is valid in the high-current limit when the precessional frequency of the molecular spin is
small compared to the electronic time scales. Unlike previous works, our approach does not follow a perturbative route either in the tunneling between leads and the molecule or in the coupling between the electronic spin and the molecular magnetic moment. Accordingly, we can render the full dependence of the parameters of the LLG equation on the state of the molecular moment as well as on the applied bias and gate voltages.

The strictly adiabatic approximation causes a mean torque exerted by
the conduction electrons, while retardation effects result in a renormalization of the precession frequency and Gilbert damping. In addition, equilibrium and non-equilibrium fluctuations of the current cause a fluctuating (Langevin) torque. We have expressed these torques in terms of the
electronic Green's functions and have related them to scattering theory, in the
latter case extending earlier work to include an applied bias voltage. We have
concluded that in general out-of-equilibrium situations the conduction electrons
can transfer energy to the localized moment by the fluctuations and,
in the presence of spin-polarized leads, via a non-conservative (spin-transfer) torque and/or negative damping.

These mechanisms allow one to use the
anisotropic magnetic molecule in an external magnetic field as a molecular
switch which can be read out via the backaction of the molecular spin on the
transport current.
When the molecule
is attached to metallic leads in a uniaxial setup, we have turned the
Langevin equation into a Fokker-Planck equation allowing us to calculate the
switching rates between the two stable spin orientations. Transitions
between these states are driven by the fluctuations which we have analyzed-- in
addition to the mean
torque, damping, and the current-- as a function of the applied gate and bias
voltages and the orientation of the molecular spin. In the presence of
spin-polarized leads, the switching dynamics is dominated by the
non-conservative (spin-transfer) part of the current-induced torque, which enables switching
between the spin orientations by reversing the direction of the electronic
current.

The above mentioned features of the dynamics of the local magnetic moment
are also common in layered spintronic devices. However, in the present case,
the different coefficients that govern the dynamics of the molecular magnetic moment show a strong
dependence on the bias voltage determined by the electronic structure of the molecule (see Fig.\ 3). The latter property
also determines the behavior of the
electronic current, where features of the dynamics of the magnetic moment take place in combination with coherent tunneling of molecular systems, as signalized for instance in the current and the differential conductance (see Fig.\ \ref{fig:current}).

We have considered a generic and standard model for the molecule which applies to a wide type of molecular systems, provided that a sufficiently large current flows through the molecule and that the magnetic moment is sufficiently large to fulfill the adiabatic condition assumed in the NEBO treatment. In particular, good candidates can be
the Mn$_{12}$- or Fe$_8$-based devices. These systems 
are described by microscopic Hamiltonians of the type we considered in this work, and have rigid magnetic cores with $M=10$, and an anisotropy barrier $D M^2$ of the order of a few meV.\cite{fried,revs} Classical descriptions of their spin dynamics have been presented for these  molecules in contact to phononic environments. \cite{zueco} The crucial
parameters in order to achieve the adiabatic regime in our setup, should be a good enough contact to the electrodes and  a sufficiently high applied bias voltage, leading to a short dwell time of the electrons in the molecule. To be more specific, we estimate for the Mn$_{12}$- or Fe$_8$ systems with a rather large magnetic anisotropy that the Born-Oppenheimer approximation can be applied when the current through the device exceeds $\sim 10 \mathrm{nA}$.

\begin{acknowledgments}
We acknowledge discussions with S.\ Viola Kusminskiy and are grateful for
financial
support through SFB 658 and an institute partnership funded through the
Alexander-von-Humboldt foundation. L.A.\ and G.S.L.\ thank
from CONICET and MINCyT (Argentina) for support. LA thanks the J.S.\ Guggenheim
Foundation for support.
\end{acknowledgments}

\appendix

\section{Green's functions, $s_l^0(t,{\bf M})$,
$\gamma_s^{lk}(t,{\bf M})$ and noise
correlator}
\label{app:coefficients}
We approximate the self-energy to be
independent of energy. In this wide band limit Eq.\ \eqref{eq:selfenergyR} becomes
\begin{align}
\Sigma_{\alpha,\sigma }^{R}(\omega)  &  \simeq \mathrm{Re}\left(\Sigma_{\alpha,\sigma }^{R}\right)-\pi
i\,\nu_{\alpha,\sigma}\left|w_{\alpha}\right|^{2}.\label{eq:sigmarapp}
\end{align}
with the approximately constant density of states
$\nu_{\alpha,\sigma}(\omega)\simeq \nu_{\sigma}$ and
$\left|w\right|^{2}=\left(\left|w_{L}\right|^{2}+\left|w_{R}\right|^{2}
\right)/2$. We introduce the abbreviations
\ba 
\epsilon_{\sigma}&=&\epsilon_{0}+\sum_{\alpha}\mathrm{Re}\left(\Sigma_{\alpha,
\sigma}^{R}\right)\\
\Gamma_{\sigma}/2&=&-\mathrm{Im}\left[\Sigma_{L,\sigma}^{R}(\omega)+\Sigma_{R,
\sigma}^{R}(\omega)\right]\simeq \pi\,\nu_{\sigma}\left|w\right|^{2},
\ea 
and we will use the notation
$\Sigma^R_{c,s}=
[\Sigma^R_{\uparrow} \pm \Sigma^R_{\downarrow}]/2$ and $\Gamma_{c,s}=
\sum_{\alpha} \Gamma_{\alpha,c(s)}$, with $\Gamma_{\alpha,c(s)}=(\Gamma_{\alpha,
\uparrow} \pm \Gamma_{\alpha, \downarrow})/2$, taking into account possibly
spin-polarized leads.

From Eq.\ \eqref{eq:G0Rinv} we find for the frozen retarded Green's
function
\begin{align}
G_{0}^{R}(\bf{M},\omega) & =
\frac{1}{\tilde{\omega}^{2}-\tilde{b}^{2}}\,\left[\tilde{\omega}\,
1+\mathbf{\tilde{b}}
\cdot\boldsymbol{\sigma}\right]\nonumber\\
 & =
\frac{1}{2}\frac{1+\boldsymbol{\sigma}\cdot\mathbf{\tilde{b}}/\tilde{b}}{\tilde{
\omega } -\tilde{b}}
+\frac{1}{2}\frac{1-\boldsymbol{\sigma}\cdot\mathbf{\tilde{b}}/\tilde{b}}{\tilde
{\omega
}+\tilde{b}},\label{eq:G0Rapp}
\end{align}
with $\tilde{\omega}=\left(\omega-\epsilon_{0}-\Sigma^{R}_c\right) \simeq
\omega-\epsilon+i\Gamma_c/2$. Here, we include the antisymmetric part of the self-energy
in the effective magnetic field,
\begin{align}
\mathbf{\tilde{b}}(t)=\frac{1}{2}(J\mathbf{M}(t)+ g_e
\mathbf{B}) +\Sigma^R_s \hat{\mathbf{e}}_z.
\end{align}
After some algebra we find the following expression for the lesser
Green's function \eqref{eq:Gless}:
\begin{align}
G^<_0 (\mathbf{M},\omega)=&G_I^< (\mathbf{M},\omega) + G_b^<
(\mathbf{M},\omega) {\bf b} \cdot \boldsymbol{\sigma} + 
G_z^< (\mathbf{M},\omega) \sigma^z \nonumber\\
&+ G_t^< (\mathbf{M},\omega)
\boldsymbol{\sigma} \cdot (\hat{\mathbf{e}}_z \times {\bf b}),\label{eq:G0<app}
\end{align}
where the coefficients are given by
\begin{align} \label{ges}
G_I^< (\bf{M},\omega) =& \frac{1}{|\Delta(\mathbf{M},\omega)|^2} \{ 
\Sigma_c^<(\omega) 
[|\tilde{\omega}|^2+|\mathbf{\tilde{b}}|^2]\nonumber\\
  &+ \Sigma_s^<(\omega) \,2
[\mbox{Re}[\tilde{\omega}]  b_z - \frac{\Gamma_c \Gamma_s}{4}] \}
,\nonumber \\
G_b^< (\bf{M},\omega) =& \frac{2}{|\Delta(\mathbf{M},\omega)|^2} \{
 \Sigma_c^<(\omega) \mbox{Re}[\tilde{\omega}] +  \Sigma_s^<(\omega) b_z \}
,\nonumber \\
G_z^< (\bf{M},\omega) =& \frac{1}{|\Delta(\mathbf{M},\omega)|^2}
\{- \Sigma_c^<(\omega) \frac{\Gamma_c \Gamma_s}{2} \nonumber\\
  &+\Sigma_s^<(\omega)[ |\tilde{\omega}|^2 - |\mathbf{\tilde{b}}|^2 +
\frac{\Gamma_s^2}{2}]\} 
,\nonumber \\
G_t^< (\bf{M},\omega) =&
\frac{1}{|\Delta(\mathbf{M},\omega)|^2}\{\Sigma_c^<(\omega) \Gamma_s -
\Sigma_s^<(\omega) \Gamma_c\}.
\end{align}
We use $\Delta(\mathbf{M},\omega)=\tilde{\omega}^{2}-\tilde{b}^{2}$ and $\mathbf{\tilde{b}}\simeq \mathbf{b} - i (\Gamma_{s}/2)\hat{\mathbf{e}}_z$.
Substituting the above expressions for $\Gamma_{c,s}$, it can be seen
that $G_t^<(\mathbf{M},\omega)=0$, for $\Gamma_{L,\sigma}= \Gamma_{R,\sigma}$, implying that this component of the Green's function contributes only for polarized
leads. Note that the corresponding expressions for the
larger Green's function $G^>_0(\mathbf{M},\omega)$ are obtained by replacing
$\Sigma^<_{c,s}$ by  $\Sigma^>_{c,s}= -i
\sum_{\alpha}\left(1-f_{\alpha}(\omega)\right) \Gamma_{\alpha,c(s)}$ in the
expressions above.

Using the Green's functions expressions, 
we find for the mean value of the electronic spin at the molecule
\begin{align} 
\mathbf{s}^0({\bf M}) =  \int \frac{d \omega}{2 \pi i} 
\{G^<_b \mathbf{b} +
G^<_z\hat{\mathbf{e}}_z+
G^<_t (\hat{\mathbf{e}}_z \times {\bf b}) \},\label{eq:s0}
\end{align}
resulting in Eq.\ \eqref{eq:s0uniaxial} in the case of axial symmetry. The explicit
expression for the component parallel to $\mathbf{M}$ reads
\begin{align}
  s_M ({\bf M})  & = - \frac{i J}{M} \int \frac{d \omega}{2 \pi}
G^<_b ({\bf M}, \omega).
\end{align}

The correction due to retardation effects
associated with the slow dynamics of the molecular spin are captured by the
matrix $\gamma^{lk}$, see Eq.\ \eqref{eq:gamma}. The symmetric part of this matrix,
\begin{align}\label{eq:gammasGr}
  \gamma_s^{lk}(\mathbf{M}) =  \frac{J^2}{4} \int \frac{d\omega}{2\pi}
\mbox{tr}[\sigma^l
\partial_{\omega} G_0^> \sigma^k G_0^<]_s,
\end{align}
describes Gilbert damping of the molecular spin, induced by the coupling to the
electrons.
The antisymmetric part of the matrix $\gamma$ is given by
\begin{align}\label{eq:gammaaGr}
 \gamma^{lk}_a(\mathbf{M}) &=  \frac{J^2}{4}  \int \frac{d\omega}{2\pi}
\mbox{tr}[\sigma^l
\partial_{\omega}\left(G_0^R + G_0^A\right) \sigma^k G_0^<]_a.
\end{align}

Considering a setup with unpolarized leads and the external magnetic field pointing along the anisotropy axis, hence $\mathbf{b} = \left(J \mathbf{M} + g_e B \hat{\mathbf{e}}_z\right)/2$, Eq.\ \eqref{eq:gammasGr} becomes
\begin{align}\label{eq:gammasN}
  \gamma^{kl}_s(\mathbf{M}) =& \frac{J^2}{4} \int \frac{d \omega}{2 \pi}
\sum_{\alpha \beta}  
\frac{\Gamma_\alpha \Gamma_\beta (-\partial_\omega
f_\beta)}{\prod_{\pm}\left[(\omega-\epsilon\pm
b)^2+(\Gamma/2)^2\right]}\delta_{kl} \nonumber\\
     +&\frac{J^2}{2} \int \frac{d \omega}{2 \pi} \sum_{\alpha \beta}  
\frac{\Gamma_\alpha \Gamma_\beta (-\partial_\omega
f_\beta)\, (\omega-\epsilon)^2}{\prod_{\pm}\left[
(\omega-\epsilon \pm b)^2+(\Gamma/2)^2\right]^2} b_k b_l.
\end{align}
This will be decomposed into a term proportional to the unit matrix and a
projector onto the $z$-axis, as described in Sec.\ \ref{sec:applications}. Note that the sign
of the eigenvalues of $\gamma^{kl}_s$ is fixed, corresponding
to \textit{damping} in and out-of equilibrium.
As described in the main text, the prefactor $\alpha$ in Eq.\ \eqref{eq:LLG2} is
given by
$\alpha(\mathbf{M}) =1/\left(1 +
\mathbf{C}\cdot \mathbf{M}\right)$, with $\mathbf{C}$ defined in Eq.\ \eqref{eq:C}. This
becomes
\begin{align}\label{eq:CN}
  \mathbf{C}(\mathbf{M}) =  \int \frac{d \omega}{2 \pi}
\sum_{\alpha}
\frac{J^2 \Gamma^2\,
\Gamma_\alpha f_\alpha\, (\omega-\epsilon)}{\prod_{\pm}\left[(\omega-\epsilon
\pm b)^2+(\Gamma/2)^2\right]^2} \mathbf{b},
\end{align}
where  we have inserted $G_0^<$, Eq.\ \eqref{eq:G0<app}, and the corresponding expression for $G_0^>$ into Eq.\ \eqref{eq:gammaaGr}.

\section{Fokker-Planck equation}\label{app:fp-derivation}
In this appendix we derive the Fokker-Planck equation from the Langevin
equation and obtain an expression for the mean first passage time, following
standard arguments.\cite{zwanzig-2001}

We note that the
probability distribution for the molecular spin is conserved for all $t$, $\int\mathrm{d}\mathbf{M}\,
f(\mathbf{M},t)=1$. 
Hence, we can write a continuity equation for the probability distribution,
\begin{equation}
\partial_{t}f(\mathbf{M},t)+\partial_{\mathbf{M}}
\cdot\left(\dot{\mathbf{M}} \,f(\mathbf{M},t)\right)=0.
\end{equation}
Inserting Eq.\ (\ref{eq:LLG2}) for $\dot{\mathbf{M}}$ we get
\begin{equation}
\partial_{t}f(\mathbf{M},t)=-L\, f(\mathbf{M},t)-\partial_{\mathbf{M}}
\cdot\left(\alpha
\boldsymbol{\xi}(t)f(\mathbf{M},t)\right)\label{eq:FPderivation2} ,
\end{equation}
where $\boldsymbol{\xi}(t) = {\bf M}\times \boldsymbol{\delta{
\cal B}}$ and the
differential operator $L$ is defined via its action on the function
$f(\mathbf{M},t)$ as
\begin{align}
L\, f&= \partial_{\mathbf{M}}\cdot \left(\alpha  {\bf M} \times
\left[-\partial_{\mathbf{M}} U - J \mathbf{s}^0  - \gamma_s \dot{\mathbf{M}} 
\right] f\right).
\end{align}
From this follows the implicit solution
\begin{align}
f(\mathbf{M},t)=&e^{-tL}f(\mathbf{M},0)\nonumber\\&-\int_{0}^{t}\mathrm{d}t'\,
e^{-(t-t')L}\partial_{\mathbf{M}}\cdot\left(\boldsymbol{\xi}(t')\,
f(\mathbf{M},t')\right).
\end{align}
 Inserting this again in Eq.\ \eqref{eq:FPderivation2} and averaging over noise,
denoted by $P(\mathbf{M},t)=\langle f(\mathbf{M},t)\rangle$, yields the
Fokker-Planck equation
\begin{align}
\partial_{t}P(\mathbf{M},t) &=-L\,P(\mathbf{M},t)
+\frac{1}{2}\partial_{\mathbf{M}}\cdot(\alpha^2
\mathbf{\tilde{D}})\cdot\partial_{\mathbf{M}}
P(\mathbf{M},t)\nonumber \\
  &= F\, P(\mathbf{M},t)\label{eq:fp1},
\end{align}
where we use that the noise is Gaussian and delta-function correlated,
$\langle\xi_k(t) \xi_l(t')\rangle = \tilde{D}^{kl} \delta(t-t')$ and introduce
the Fokker-Planck operator $F$.

We consider the distribution $P(\mathbf{M},t)$ of $\mathbf{M}$ which have been
at $\mathbf{M}_0$ at time $t=0$ and are inside a given volume at time $t$. The
mean first passage time is then given by
\begin{align}
  \tau(\mathbf{M}_0) = \int dt\, t \int d\mathbf{M}
\frac{-dP(\mathbf{M},t)}{dt},\label{eq:tauderiv01}
\end{align}
where $-\int d\mathbf{M}\frac{dP(\mathbf{M},t)}{dt}$ is the distribution of
first passage times and $\int d\mathbf{M} P(\mathbf{M},t)$ gives the number of
$\mathbf{M}$ which are still in the volume of consideration at time $t$. The
distribution of $\mathbf{M}$ is $P(\mathbf{M},t)= e^{t
F}\delta(\mathbf{M}-\mathbf{M}_0)$ with $P(\mathbf{M},t)=0$ when $\mathbf{M}$ is
at the boundary of the volume. We insert this into Eq.\ \eqref{eq:tauderiv01} so
that after integration by parts
\begin{align}
\tau(\mathbf{M}_0)= \int dt \int \mathbf{M} e^{t
F}\delta(\mathbf{M}-\mathbf{M}_0) = \int dt e^{t F^\dagger} 1,
\end{align}
with the adjoint Fokker Planck operator $F^\dagger$. This results in the
differential equation
\begin{align}
  F^\dagger \tau(\mathbf{M}) = -1
\end{align}
for the mean first passage time with an absorbing boundary condition.

\end{document}